\documentclass{article}

\usepackage{arxiv}

\usepackage[T1]{fontenc}    
\usepackage{hyperref}       
\usepackage{url}            
\usepackage{booktabs}       
\usepackage{amsfonts}       
\usepackage{nicefrac}       
\usepackage{microtype}      
\usepackage{amsmath}
\usepackage{amssymb}
\usepackage{dsfont}
\usepackage{amsmath, amsthm}
\usepackage{graphicx}
\usepackage{upgreek}
\usepackage[colorinlistoftodos]{todonotes}
\usepackage{datetime}
\usepackage{enumitem}
\theoremstyle{definition}
\newtheorem{definition}{Definition}[section]
\newtheorem{theorem}{Theorem}[section]
\newtheorem{lemma}[theorem]{Lemma}
\newtheorem{proposition}[theorem]{Proposition}

\newtheorem{remark}[theorem]{Remark}

\date{}
\title{A Characterization Framework for Stable Sets and Their Variants}

\author{
  Athanasios Andrikopoulos\thanks{Professor  (https://www.ceid.upatras.gr/webpages/faculty/aandriko/)} \\
  Dept. of Computer Engineering and Informatics\\
  University of Patras\\
  Patras, 26504, Greece \\
  \texttt{aandriko@ceid.upatras.gr} \\
\And
Nikolaos Sampanis\\
 Dept. of Computer Engineering and Informatics\\
 University of Patras\\
 Patras, 26504, Greece \\
 \texttt{nsampanis@upatras.gr} \\
}

\begin{document}.
\maketitle

\begin{abstract}
The theory of optimal choice sets offers a well-established solution framework in social choice and game theory. 
In social choice theory, decision-making is typically modeled as a maximization problem.
However, when preferences are cyclic -- as can occur in economic processes -- the set of maximal elements may be empty, raising the key question of what should be considered a valid choice.
To address this issue, several approaches -- collectively known as general solution theories -- have been proposed for constructing non-empty choice sets. Among the most prominent in the context of a finite set of alternatives are the Stable Set (also known as the Von Neumann-Morgenstern set) and its extensions, such as the Extended Stable Set, the socially stable set, and the $m$-, and $w$-stable sets.
In this paper, we extend the classical concept of the stable set and its major variants - specifically, the extended stable set, the socially stable set, and the $m$- and $w$-stable sets - within the framework of irreflexive binary relations over infinite sets of alternatives. Additionally, we provide a topological characterization for the existence of such general solutions.
\end{abstract}


\section{Introduction} 
The classical rationality conditions in choice theory formalize the thesis that rational choice involves selecting an alternative in such a way that no other alternative would be considered better or more preferable.
That is, each individual makes choices by selecting, from each 
feasible set
of alternatives, those which maximize his own preference relation.
According to this hypothesis, the set of choices from a given set of alternatives in which a dominance relation 
is defined consists of the set of maximal elements with respect to this dominance relation, which is known as the {\it core}.
The ordered pair $(X,R)$, where $X$ is a (finite or infinite)
non-empty set of alternatives 
and $R$ is a dominance relation over $X$, is called an {\it abstract decision problem}. 
The set of maximal elements (i.e., the core) of an abstract decision problem is often 
empty. In such cases, it is important to specify criteria that yield reasonable sets of alternatives as solutions. 
In the choice and game theories, a number of theories -- known as {\it general solution theories} -- have been proposed to 
take over the role of maximality 
in the absence of maximal elements. 
Any solution that includes the set of maximal alternatives 
is called {\it core-inclusive}.
Because general solution theories generalize the notion of core, a logical requirement is that they be core-inclusive.
One of the most important general solution concepts is the Schwartz set,
which is equivalent 
to the admissible set by Kalai and Schmeidler \cite{KS} or to the dynamic solutions concept of Shenoy \cite{she} in game theory.
The Schwartz set is not only non-empty for every finite abstract decision problem, but also core-inclusive. 
However, this solution presents a drawback:
it may include all the alternatives under consideration. A different approach to finding general solution concepts for solving choice problems, which is still being studied 
and improved today is the concept of stable set introduced by Von Neumann and Morgenstern \cite{von}.
The stable sets solution is core-inclusive and behaves well in acyclic dominance relations. However, 
the theory of stable sets has a significant flaw in that it can be empty in the case of odd cycles.
A wide range of general solution concepts of the Schwartz and Von Neumann and Morgenstern type
has been developed to address the shortcomings of the core in abstract decision problems, particularly when maximal elements fail to exist.
 Among the most notable are the generalized stable set \cite{han}, the extended stable set \cite{HDS}, 
 the socially stable set \cite{DM}, the $m$-stable set \cite{per}, and the $w$-stable set \cite{han}.
These concepts, which differ in their definitions of internal and external stability and apply to a wide range of abstract decision problems, have been proposed in order to:
(1) resolve the problem of solution existence,
(2) extend the notion of maximal alternative sets, and
(3) define stability in a way that prevents any chosen alternative from being dominated by another, embedding this stability within the structure of the solution itself (see \cite{KPS}, \cite{KS}, \cite{nic}, \cite{per}, \cite{sch}).

Topology provides a powerful mathematical framework for analyzing complex systems in economics and the social sciences. Two fundamental topological concepts -- continuity and compactness -- play a critical role in ensuring the existence and stability of equilibria and optimal choices. Continuity guarantees smooth behavior in preferences, while compactness ensures the existence of solutions within closed and bounded domains. These properties underpin foundational results in general equilibrium theory, game theory, and decision theory.
In the context of abstract choice problems, a common approach to establishing the existence of non-empty choice sets over infinite sets of alternatives - where an arbitrary dominance relation is defined - is to assume topological properties such as compactness and continuity.
Building on this foundation,  Andrikopoulos proposed a topological characterization of the existence of the Schwartz set and the generalized stable set in infinite sets of alternatives in \cite{and}, \cite{and77} and \cite{and1}, \cite{and7}, respectively.

In this paper, we provide characterizations of the 
 existence of certain notions of stable sets solutions -- namely, 
the Stable Set, Extended Stable Set, Socially Stable Set, $m$-Stable Set and $w$-Stable Set -- 
 for irreflexive binary relations over non-finite sets of alternatives.
 Beyond the concept of compactness, which we use to establish the basic existence theorems, the other key notions we rely on are those of continuity, as defined by Andrikopoulos in \cite{and1}, \cite{and7}, and the notion of contraction relation, as defined in \cite{BCM}.

\section{Notations and definitions}
An abstract decision problem is divided into two parts. One is an arbitrary set $X$ of alternatives (called the {\it ground set}) 
from which an individual or group must select. In most cases, there are at least two alternatives to choose from. Otherwise, 
there is no need to make a decision. The other is a dominance relation over this set, which reflects preferences or evaluations 
for different alternatives.
Preferences or evaluations over $X$ are modeled by a binary relation $R$. We sometimes abbreviate $(x,y)\in R$ as $xRy$. 
Several structural properties of the relation $R$ are particularly relevant. 
The {\it diagonal relation} $\Delta \subseteq X \times X$ is defined as
$
\Delta= \{ (x, x) \mid x \in X \}.
$
The relation is said to be {\it reflexive} if every element is related to itself; that is, $(x,x)\in R$ for all $x \in X$. If no element is related to itself -i.e., 
$(x,x)\notin R$ for all $x \in X$ - then $R$ is called {\it irreflexive}. Moreover, $R$ is called {\it transitive} 
if for any $x, y, z \in X$, the relations $(x,z) \in R$ and $(z,y) \in R$ imply $(x,y) \in R$, and {\it antisymmetric} if for all 
$x, y \in X$, $(x, y) \in R$ and $(y, x) \in R$ imply $x = y$.
A {\it partially ordered set} ({\it poset}) is a pair $(X,\preceq)$, where $X$ is a set and $\preceq$ is a reflexive, transitive and antisymmetric binary relation on $X$. 
An element $x\in X$ is a {\it lower bound} of a subset $A\subseteq X$ if $aRx$ for all $a \in A$, while $y\in X$ is an {\it upper bound} of 
$A$ if $yRa$ for all $a \in A$.
The {\it supremum} of a subset $A$ of an ordered set is the smallest of all upper bounds of $A$, if such an element exists.
A subset $D \subseteq X$ is called {\it directed} if it is non-empty and every finite subset of $D$ has an upper bound in $D$.
For every directed set $D \subseteq X$, if the supremum of $D$ exists, it is denoted by $\bigvee D$.
The {\it join} of two elements $x$ and $y$, denoted $x\vee y$, is the least element greater than or equal to both $x$ and $y$.  
The {\it meet} of $x$ and $y$, denoted $x\wedge y$, is the greatest element less than or equal to both.
A {\it lattice} is an ordered set $(X, R)$ such that for every pair of elements $x, y \in X$, both the join and the meet exist.  
The structure $(X, R)$ is called a {\it complete lattice} if all joins and meets exist for arbitrary subsets of $X$.

We exclusively focus on abstract decision problems with a non-empty arbitrary set $X$ and an irreflexive dominance relation $R$ on $X$.
When representing abstract decision problems, the pair $(X,R)$ is used. 
We denote by $\Omega(X)$ the set of abstract decision problems on $X$.
For any \( x \in X \), the sets
$Rx = \{ y \in X \mid yRx \}$ and $xR = \{ y \in X \mid xRy \}$
are called the \emph{upper contour set} and the \emph{lower contour set} of \( R \) at \( x \), respectively.
The asymmetric part of $R$ is defined by 
$P(R)=\{ (x,y) \in X\times X \vert (x,y)\in R$ and $(y,x)\notin R \}$.
The {\it transitive closure} of $R$ is the relation $\overline{R}$ defined as follows:
For all $x, y\in X$, $(x,y)\in \overline{R}$ if and only if there exist $K\in \mathbb{N}$ and $x_{_0},...,x_{_K}\in X$ such that
$x=x_{_0}, (x_{_{k-1}},x_{_k})\in R$ for all $k\in \{1,...,K\}$ and $x_{_K}=y$.
A subset $D\subseteq X$ is $R$-{\it undominated} if and only if for no $x\in D$ is there a $y\in X\setminus D$ such that $yRx$. 
A subset $Y\subseteq X$ is an $R$-{\it cycle} if for all $x,y\in Y$, we have 
$(x,y)\in \overline{R}$ and $(y,x)\in \overline{R}$. A {\it Top $R$-cycle} is an $R$-cycle 
which is maximal with respect to set-inclusion.
We say that $R$ is {\it acyclic} if there does not exist an $R$-cycle. 
An alternative $x\in X$ is $R$-{\it maximal} with respect to a binary relation $R$,
if $(y,x)\in P(R)$ for no $y\in X$. $\mathcal{M}(X,R)$ denotes the elements of $X$ that are $R$-maximal in $X$, hence, 
$\mathcal{M}(X,R)=\{x\in X\vert$ for all $y\in X, yRx$ implies $xRy\}$.
In what follows, $\mathfrak{P}(X)$ denotes the family of non-empty subsets of $X$.
A choice function $\mathcal{C}$ is a mapping that assigns to each $A\in \mathfrak{P}(X)$ a subset of $A$: 
$\mathcal{C}\rightarrow 2^X$ such that for all $A\in \mathfrak{P}(X)$, $\mathcal{C}(A)\subseteq A$.
The traditional choice-theoretic approach takes behavior as rational if there is a binary relation $R$ such that for 
each non-empty subset $A$ of $X$, $\mathcal{C}(A)=\mathcal{M}(A,R)$ ($\mathcal{M}(A,R)$ denotes the elements $X$ that are $R$-maximal in $A$). To deal with the case where the set 
of maximal elements is empty, Schwartz in \cite[p. 142]{sch} has proposed the general solution concept known as
{\it Generalized Optimal-Choice Axiom} ($\mathcal{G}\mathcal{O}\mathcal{C}\mathcal{H}\mathcal{A}$): 
For each $A\subseteq X$, $\mathcal{C}(A)$ is equivalent to the union of all minimal 
$R$-undominated subsets of $A$.
From now on we will denote the union 
of all minimal 
$R$-undominated subsets of an abstract decision problem $(X,R)$ by $\mathcal{S}_{ch}(X,R)$
and call it the {\it Schwartz set}.
Deb in \cite{deb} shows that $\mathcal{S}_{ch}(X,R)=\mathcal{M}(X,\overline{P(R)})$ (see also \cite{and3}).
According to the generalization of Deb's Theorem in \cite{and3} and 
\cite[Theorem 19]{and4},  $\mathcal{S}_{ch}$ is equivalent to the union of all $P(R)$-undominated 
elements and all top $P(R)$-cycles in $X$. It
 is also equivalent to the notion of admissible set in game theory defined by Kalai and Schmeidler in \cite{KS} and 
 the notion of dynamic solutions defined by Shenoy in \cite{she}.
We say $x$ {\it traps} $y$, written $x\,\mathcal{T}\,y$, if $xP(R)y$ and $(y,x)\notin \overline{P(R)}$ (see \cite[p. 499]{dug}).
The {\it untrapped set} of $R$, written $\mathcal{U}\mathcal{T}(X,R)$ is the set of 
$\mathcal{T}$-maximal elements: $\mathcal{U}\mathcal{T}(X,R) = \mathcal{M}(X,\mathcal{T})$.
Following \cite[section~4]{and4}, we will henceforth refer to this set as the {\it Duggan set}.
By Theorem 3 in \cite{dug} we have that $\mathcal{S}_{ch}(X,R)\subseteq \mathcal{U}\mathcal{T}(X,R)$.
With the aim of the aforementioned notions, we proceed to generalize the following classical concepts (see \cite{han}, \cite{van} and \cite{von}), which will be used throughout this paper, for arbitrary binary relations. Specifically, a subset $V$ of $(X,R)$ is called a {\it Von Neumann-Morgenstern stable set}  if (i)
no alternative in $V$ is dominated with respect to $P(R)$ by another alternative in
$V$, and (ii) any alternative outside $V$ is dominated with respect to $P(R)$ by an alternative inside $V$.
The first
property is called {\it internal stability of stable sets} and the second property {\it external stability of stable sets}. 
Denote by $\mathcal{S}(X,R)$ the collection of stable sets of $(X,R)$.
 A set $V\subseteq X$ is called a {\it generalized stable set} if it is a stable set with respect to 
 $\overline{P(R)}$.
Similarly, as in the case of stable sets, we have the notions of
 {\it internal stability of generalized stable sets} and {\it external stability of generalized stable sets}, respectively. 
 Denote by $\mathcal{GS}(X,R)$ the collection of generalized stable sets of $(X,R)$. 
 Let $R|_{_T}$ be the sub-relation of $R$ with the restriction to $T\subseteq X$. 
 A set $V\subseteq X$ is called a {\it socially stable set} of $(X,R)$ if:
 (i) $\forall x, y\in V$, if $x\overline{P(R)}|_{_V}y$, then $y\overline{P(R)}|_{_V}x$;
 (ii) $\forall y\in X\setminus V$, there is an $x\in V$ such that $(x,y)\in P(R)$.
 Denote by $\mathcal{SS}(X,R)$ the collection of socially stable sets of $(X,R)$. 
 Condition (i) is called {\it internal stability of socially stable set} and condition (ii) is called {\it external stability of socially stable set}.
 A set $V\subseteq X$ is called an {\it $m$-stable set} of $(X,R)$ if:
 (i) $\forall x, y\in V$, if $x\overline{P(R)}y$, then $y\overline{P(R)}x$;
 (ii) $\forall x\in V$, there is no 
 $y\in X\setminus V$ such that
 $(y,x)\in \overline{P(R)}$.
 Denote by $\mathcal{MS}(X,R)$ the collection of $m$-stable sets of $(X,R)$. 
 Condition (i) is called {\it internal stability of $m$-stable set} and condition (ii) is called {\it external stability of $m$-stable set}.
 A set $V\subseteq X$ is called a $w$-{\it stable set} of $(X,R)$ if
(i) $\forall x,y \in V$, $(x,y)\notin\overline{P(R)}$;
(ii) $\forall x \in V$ and $y\in X\setminus V$, if $y\overline{P(R)}x$ then $x\overline{P(R)}y$.
Condition (i) is called {\it internal stability of $w$-stable set} and condition (ii) is called {\it external stability of $w$-stable set}.
 Denote by $\mathcal{WS}(X,R)$ the collection of $w$-stable sets of $(X,R)$.

An abstract decision problem $(X,R)$ is called {\it strongly connected} if $x\overline{P(R)}y$ for all $x, y\in X$. 
A {\it strong component} of an abstract decision problem $(X,R)$
is an abstract decision problem $(Y,R|_{_Y})$, $Y\subseteq X$, satisfying the following properties:
($\mathfrak{i}$) $(Y,R|_{_Y})$ is strongly connected;
($\mathfrak{i}\mathfrak{i}$) no abstract decision problem $(Y^{\prime},R|_{_{Y^{\prime}}})$ with $Y^{\prime}\supset Y$ 
is strongly connected.
Note that when an element $x$ is not on any $P(R)$-cycle, it forms a singleton strongly connected component $\{x\}$ by itself.
Clearly, the set of strongly connected components forms a partition of the space $(X,R)$.
The
{\it contraction} of $(X, R)$ is an abstract decision problem $(\Xi,\widetilde{R})$ where
\par
1. $\Xi=\{X_{_i}\vert i\in I\}$ is the collection of ground sets of the strong components of $(X,R)$;
\par
2. for any $X_{_i}, X_{_j}\in \Xi$, $X_{_i} \widetilde{R} X_{_j}$ if there are $x\in X_{_i}, y\in X_{_j}$ with 
$xP(R)y$. Clearly, $\widetilde{R}$ is acyclic by definition.

In what follows, $\mu(\Xi,\widetilde{R})=\{X^{\ast}_i\vert i\in I\}$ denotes the family of ground sets which 
are $\widetilde{R}$-maximal in $\Xi$.

For any $x,y \in X$, we say $x$ and $y$ are {\it equipotent}, denoted by $x\mathfrak{I}y$, if either $x=y$ or 
$x\overline{P(R)}y$ and $y\overline{P(R)}x$.
Let $(X,R)\in \Omega(X)$ and $x, y\in X$. We say $x$ $\widetilde{\omega}$-dominates $y$, denoted by $xR^{^{\widetilde{\omega}}}y$, if there are
$z, w\in X$ such that $x\mathfrak{I}z$, $zP(R)w$ and $w\mathfrak{I}y$ (see \cite[Definition 1]{HDS}).
 A set $V\subseteq X$ is called an {\it extended stable set} of $(X,R)\in \Omega(X)$ if (i) $\forall x,y \in V$, 
 $(x,y)\notin R^{^{\widetilde{\omega}}}$;
 (ii) $\forall y \in X\setminus V$, there is an $x\in V$ such that $xR^{^{\widetilde{\omega}}}y$.
Condition (i) is called {\it internal stability of extended stable set} and condition (ii) is called {\it external stability of extended stable set}.

We say that a topological space $(X,\tau)$  is {\it compact} if for 
each collection of open sets which covers $X$ there exists a finite subcollection that also covers $X$.
If $\approx$ is an equivalence relation on $X$, then the quotient
set by the relation $\approx$ will be denoted by ${\mathfrak{X}}^\approx$ , and its elements (equivalence
classes) by $X^{\approx}$. 
Let the projection map $\pi: X\to {\mathfrak{X}}^\approx$ which carries each point of $X$ to the element of ${\mathfrak{X}}^\approx$ that contains it. In the quotient topology induced by 
$\pi$, a subset $U$ of ${\mathfrak{X}}^\approx$ is open in ${\mathfrak{X}}^\approx$ if and only if $\pi^{-1}(U)$ is open in $X$. Thus, the typical open set in ${\mathfrak{X}}^\approx$ is a collection of equivalence classes whose union is an open set in $X$ . The quotient topology associated with a topological space $(X,\tau)$ and an equivalence relation $\approx$ will be denoted by $\tau\!\!_{_{{\mathfrak{X}}^\approx}}$.

In the following, for the sake of uniformity and simplification of the proofs, if the equivalence relation arises from the concept of strong connectedness, 
then we denote ${\mathfrak{X}}^\approx$ by $\Xi$ and $X^{\approx}$ by $X^{\ast}$.

\section{Characterization of a stable set and its variants} \vspace{-0.2cm}

We will start with a result that will form the basis for some of the results that follow. 
Given an abstract system $(X, R)$, define the poset 
\[
\mathcal{P} = (X, \preceq_{_R}),
\]
where $\preceq_{_R} = P(\overline{P(R)}) \cup \Delta$, with $P(R)$ denoting the asymmetric part of $R$ and $\Delta$ the diagonal relation on $X$. Clearly, if $R$ is an acyclic binary relation, then $\preceq_{_R} = \overline{R} \cup \Delta$.

\begin{definition}
Let \( (X, \tau) \) be a topological space and \( R \subseteq X \times X \) a binary relation.
We say that $(X, \tau)$ has the \emph{$T_1$-order separation property with respect to} $R$ if for any $x, y \in X$ with $y \succ_{_R} x$:
\begin{itemize}
    \item[(i)]
        there exists an open neighborhood $U$ of $x$ such that $y \notin U$;
    \item[(ii)]
        for every net $(x_a)_{a \in A}$ converging to $x$ in $\tau$, there exists an index $a_0$ such that $y \succ_{_R} x_a$ for some $a \geq a_0$.
\end{itemize}
If only condition (i) is satisfied, we say that $(X, \tau)$ has the \emph{weak $T_1$-order separation property with respect to} $R$.
\end{definition}

\begin{lemma}\label{a221}{\rm Let $(X,R)$ be an abstract decision problem, and let $\tau$ be a compact topology in $X$.
Suppose that 
$(X, \tau)$ has the weak $T_1$-order separation property with respect to $R$.
Then, the family $\mu(\Xi,\widetilde{R})$ of ground sets, which 
are $\widetilde{R}$-maximal in $\Xi$, is non-empty.
}
\end{lemma}
\begin{proof} Let $x\in X$.
If $x$ is an $\overline{P(R)}$-maximal element, then $\{x\}$ belongs to a top $P(R)$-cycle (Schwartz set).
Hence, $\{x\}\in \mu(\Xi,\widetilde{R})$.
Otherwise, there exists $y\in X$ such that $y\overline{P(R)}x$.
Similarly, if $y$ is an $\overline{P(R)}$-maximal element, then $\{y\}\in \mu(\Xi,\widetilde{R})$.
Otherwise, there exists $y_{_1}\in X$ such that $y_{_1}\overline{P(R)}y\overline{P(R)}x$.
Put
\begin{center}
$A_{_{x}}=\{y\in X\vert \emptyset\subset\overline{P(R)}y\subseteq \overline{P(R)}x\}$.
\end{center}
Since $y_1\overline{P(R)}y\overline{P(R)}x$
we conclude that $A_{_{x}}\neq \emptyset$.

We now show that $A_{_{x}}$ is closed with respect to $\tau$. 
Suppose that $t$ belongs to the closure of
$ A_{_{x}}$. Then, there exists a net $(t_{_k})_{_{k\in K}}$ in
$A_{_{x}}$ with $t_{_k}\to t$. We have to show that $t\in A_{_{x}}$, i.e., $\overline{P(R)}t\subseteq \overline{P(R)}x$.
Take any $z\in\overline{P(R)}t$. Since 
 $(X, \tau)$ has the \emph{$T_1$-order separation property with respect to} $R$,
there exists 
$k_{_0}\in K$ such that $z\overline{P(R)}t_{_{k_{_0}}}$ holds.
Hence, $z\in \overline{P(R)}t_{_{k_{_0}}}\subseteq \overline{P(R)}x$ ($t_{_{k_{_0}}}\in  A_{_{x}}$).
It follows that $\overline{P(R)}t\subseteq \overline{P(R)}x$ which implies that $t\in A_{_{x}}$.
Therefore, $A_{_{x}}$ is a closed subset of $X$.
If there exists $t^{\ast}\in A_{_{x}}$ which is $\overline{P(R)}$-maximal in $X$, then $t^{\ast}$ belongs to a top $P(R)$-cycle and thus
$\mu(\Xi,\widetilde{R})$ is non-empty. Otherwise,
for each $t\in A_{_{x}}$ there exists $y\in X$ such that
$(y,t)\in \overline{P(R)}$. It follows that $y\in A_{_x}$ ($\overline{P(R)}y\subseteq \overline{P(R)}x$). 
Given that for every $t \in A_{x}$ there exists $y_t \in A_{x}$ with $y_t \succ_{R} t$, and since $(X, \tau)$ possesses the $T_1$-order separation property relative to $R$, for each such $t$ there exists an open neighborhood $U_{y_t}$ of $t$ such that $y_t \notin U_{y_t}$.
Therefore, for each $t\in A_{_{x}}$,
the sets $U_{y_{_t}}\bigcap A_{_{x}}$
are open neighbourhoods of $t$ in the
relative topology of $A_{_{x}}$.

Hence,
\begin{center}
$A_{_{x}}=\displaystyle
\bigcup_{t\in X} (U_{y_{_t}}\bigcap A_{_{x}})$.
\end{center}
Since the space $A_{_{x}}$ is compact 
in the
relative topology, there exist $\{y_{_{t_{_1}}}, \ldots, y_{_{t_{_n}}}\}$ such that
\[
A_{_{x}} = \bigcup_{i \in \{1, \ldots, n\}} (U_{y_{_{t_i}}}\bigcap A_{_{x}}).
\]
Consider the finite set $\{y_{_{t_{_1}}}, \ldots, y_{_{t_{_n}}}\}$. 
Without loss of generality, we may assume that $i \neq j$ for all distinct $i, j \in \{1, \ldots, n\}$.
Then, for each 
$t \in A_{_{x}}=\bigcup\{U_{y_{_{t_i}}} \mid i = 1, \ldots, n\}$, 
there exists $i \in \{1, \ldots, n\}$ such that $y_{_{t_{_i}}}\succ_{_R} t$. 
Since $y_{_{t_{_1}}} \in A_{_{x}}$, it follows that 
$y_{_{t_{_i}}}\succ_{_R}y_{_{t_{_1}}}$ for some $i \in \{1, \ldots, n\}$. 
If $i=1$, then we have a contradiction. Otherwise, call this element 
$y_{_{t_{_2}}}$.
Then, we have $y_{_{t_{_2}}}\succ_{_R} y_{_{t_{_1}}}$.
Similarly, $y_{_{t_{_3}}}\succ_{_R} y_{_{t_{_2}}}\succ_{_R} y_{_{t_{_1}}}$.
As 
 $\{y_{_{t_{_1}}}, \ldots, y_{_{t_{_n}}}\}$ is finite, by an induction argument based on this logic, 
 we obtain the existence of a 
$\succ_{_R}$-cycle, that is, a $P(R)$-cycle $\widetilde{\mathcal{C}}$
which contains the elements of the set $M=\{y_{_1},y_{_2},...,y_{_n}\}$.
By the Lemma of Zorn, the family of all $P(R)$-cycles $(\widetilde{\mathcal{C}}_{_\gamma})_{_{\gamma\in \Gamma}}$,
$\widetilde{\mathcal{C}}_{_\gamma}\subseteq 
A_x$, which contain $M$ has a maximal element, which we will call 
$\widetilde{\mathcal{C}}_{_{\gamma_{_0}}}$.
We prove that $\widetilde{\mathcal{C}}_{_{\gamma_{_0}}}\in \mu(\Xi,\widetilde{R})$. 
Indeed, let $X^{\ast} \widetilde{R}\ \widetilde{\mathcal{C}}_{_{\gamma_{_0}}}$ for some $X^{\ast}\in \Xi$.
Then, there exist $t\in X^{\ast}, s\in \widetilde{\mathcal{C}}_{_{\gamma_{_0}}}$ such that 
$tP(R)s$. If for each $\lambda \in X$ we have $(\lambda,t)\notin \overline{P(R)}$, then
$X^{\ast}=\{t\}$ belongs to the Schwartz set and thus
$X^{\ast}\in\mu(\Xi,\widetilde{R})$.
Otherwise, there exists $\lambda^{\ast}\in X^{\ast}$ such that $(\lambda^{\ast},t)\in\overline{P(R)}$.
Therefore, from $(\lambda^{\ast},t)\in\overline{P(R)}$, $(t,s)\in P(R)$ and $(s,x)\in\overline{P(R)}$ we conclude that 
$\emptyset\subset\overline{P(R)}t\subseteq \overline{P(R)}x$,
which implies that
$t\in A_x$. Since $s\in \widetilde{\mathcal{C}}_{_{\gamma_{_0}}}$ we have that $(t,y_i)\in \overline{P(R)}$ for each 
$i\in \{1,2,...,n\}$. On the other hand, since $t\in A_x$ we have $(y_{i^{\ast}},t)\in \overline{P(R)}$ for some $i^{\ast}\in \{1,2,...,n\}$.
Therefore, from $(t,y_{_{i^{\ast}}})\in \overline{P(R)}$ and $(y_{_{i^{\ast}}},t)\in \overline{P(R)}$ we conclude that 
$t\in \widetilde{\mathcal{C}}_{_{\gamma_{_0}}}$, which is impossible. Hence, 
$\widetilde{\mathcal{C}}_{_{\gamma_{_0}}}\in \mu(\Xi,\widetilde{R})$. Therefore, in any case we have that $\mu(\Xi,\widetilde{R})\neq\emptyset$.
\end{proof}

We now proceed to provide a characterization of the existence of the classical stable set in the sense of von Neumann and Morgenstern  \cite{von}.

\begin{proposition}\label{pan1}
Let \( (X, R) \) be an abstract system and define the set of \( R \)-maximal elements as
\[
\mathcal{M}(X, R) = \{ x \in X \mid \{ y \in X \mid y \mathrel{R} x \} = \emptyset \}.
\]
Suppose that \( \mathcal{M}(X, R) \) is non-empty and stable with respect to \( R \).
Then:
\begin{enumerate}
    \item \( \mathcal{M}(X, R) \subseteq \mathcal{M}(X, \preceq_R) \), i.e., every \( R \)-maximal element is also maximal with respect to \( \preceq_R \);
    \item \( \mathcal{M}(X, R) \) is a stable set in the poset \( (X, \preceq_R) \).
\end{enumerate}
\end{proposition}

\begin{proof}
By assumption, \( \mathcal{M}(X, R) \neq \emptyset \), so non-emptiness holds.
We show that \( \mathcal{M}(X, R) \) is stable with respect to \( \preceq_{_R} \). Let \( x \in X \). Since \( \mathcal{M}(X, R) \) is stable under \( R \), there exists \( m \in \mathcal{M}(X, R) \) such that \( m \mathrel{R} x \).
Because \( R \subseteq \overline{R} \subseteq \preceq_{_R} \), we also have \( m \succeq_{_R} x \). Hence, for every \( x \in X \), there exists \( m \in \mathcal{M}(X, R) \) such that \( m \succeq_{_R}  x \), which shows that \( \mathcal{M}(X, R) \) is stable in \( (X, ) \).
It remains to show that elements of \( \mathcal{M}(X, R) \) remain maximal in \( (X, \succeq_{_R}) \). Suppose \( m \in \mathcal{M}(X, R) \), and assume for contradiction that there exists \( y \in X \setminus \mathcal{M}(X, R) \) such that \( y  \succeq_{_R} m \). Then \( y \mathrel{\overline{R}} m \), i.e., there exists a finite sequence \( y = x_0 \mathrel{R} x_1 \mathrel{R} \cdots \mathrel{R} x_n = m \). But this contradicts the fact that \( m \) is \( R \)-maximal.

Therefore, \( m \) has no \(  \preceq_{_R} \)-predecessor except itself, so it is maximal in \( (X, \preceq_{_R}) \). Hence,
\[
\mathcal{M}(X, R) \subseteq \mathcal{M}(X, \preceq_{_R}),
\]
and since \( \mathcal{M}(X, R) \) is stable under \( \preceq_{_R} \), the proof is complete.
\end{proof}

Let $A$ be a subset of $\mathcal{P}$. Then, 
$A^\uparrow$ and $A^\downarrow$ denote the sets of all upper and lower bounds of $A$, respectively. Let
\begin{center}
$A^\delta = (A^\uparrow)^\downarrow \quad \text{and} \quad \delta(P) = \{ A^\delta : A \subseteq P \}.$
\end{center}
$(\delta(P), \subseteq)$ is called the {\it normal completion}, or the {\it Dedekind–MacNeille completion} of $P$.

\begin{definition} Let $x, y$ be elements of an ordered set $(X, \preceq)$. We say that $x \ll y$ (read: \emph{way-below relation}) if for every directed set $D \subseteq X$ such that $\bigvee D \preceq y$, there exists $d \in D$ with $d \succeq x$.

A subset $U \subseteq X$ is called \emph{Scott-open} if:
\begin{itemize}
  \item[(i)] $U$ is an \emph{upper} (respectively, \emph{lower}) set, i.e., if $x \in U$ and $y \succeq x$ (resp.\ $x \preceq y$), then $y \in U$;
  \item[(ii)] For every directed set $D \subseteq X$, if $\bigvee D \in U$, then $D \cap U \neq \emptyset$. This condition is called \emph{inaccessible by directed joins}.
\end{itemize}

The collection of all Scott-open sets forms the \emph{Scott topology}, denoted by $\sigma$. 
The \emph{lower topology} $\omega$ on an ordered set $(X, \preceq)$ is generated by the sets of the form $X \setminus \uparrow x = \{y \in X : y \not\succeq x\}$ for any $x \in X$.
The \emph{Lawson topology} on an ordered set is defined as the supremum of the Scott topology $\sigma$ and the lower topology $\omega$:
\[
\lambda = \sigma \vee \omega.
\]
\end{definition}

\begin{lemma}\label{law}(\cite[Theorem 1-6.4]{GW}
For a complete lattice $\mathcal{L}$ the Lawson topology $\lambda(\mathcal{L})$ is a
compact $T_{_1}$-topology.
\end{lemma}

\begin{definition}\label{pan2} (\cite{fri}).
A subset \( I \) of a poset \( \mathcal{P} \) is called a \textit{Frink ideal} in \( X \) if \( Z^\delta \subseteq I \) for all finite subsets \( Z \subseteq I \). Let \( \mathrm{Fid}(X) \) denote the set of all Frink ideals.
\end{definition}

\begin{definition}\label{pan3} (\cite{ern1}).
Let \( \mathcal{P} \) be a poset and \( A, B \subseteq X \).
\begin{enumerate}
    \item We say that \( A \ll_e B \) if for all Frink ideals \( I \in \mathrm{Fid}(X) \),
    \[
    \uparrow B \cap I^\delta \neq \emptyset \quad \Rightarrow \quad \uparrow A \cap I \neq \emptyset.
    \]
    
    \item The element-wise version is defined by \( x \ll_e y \) if and only if \( \{x\} \ll_e \{y\} \). That is,
    \[
    \forall I \in \mathrm{Fid}(X), \quad y \in I^\delta \Rightarrow x \in I.
    \]
\end{enumerate}
\end{definition}

\begin{remark}\label{pan4}
This definition provides an ideal-theoretic formulation of the way-below relation using Frink ideals.

Given a Frink ideal \( I \), the condition \( y \in I^\delta \Rightarrow x \in I \) means that whenever \( y \) is an upper bound of a finite subset of \( I \), the element \( x \) must already belong to \( I \). Hence, \( x \ll_e y \) captures the idea that \( x \) is ``deep below'' \( y \) in the structure of \( X \), not merely in terms of the order \( \leq \), but in terms of ideal-theoretic approximation.
In other words, every element \( x \in X \) lies in the upper closure of the set of elements that are way-below it.
In fact,
Erné’s way-below relation is a genuine generalization of the Scott way-below relation: it extends the notion from complete lattices - where all directed suprema exist—to arbitrary posets, by replacing directed suprema with ideal-theoretic approximations. Thus, Erné’s definition coincides with the Scott relation in complete lattices, but also applies naturally to all posets.
\end{remark}

\begin{definition}\label{pan5}(\cite{ern1}).
A poset \( \mathcal{P} \) is called {\it precontinuous} if for all \( x \in \mathcal{P} \),
$
x \in \left( \{ y \in \mathcal{P} \mid y \ll_e x \} \right)^\delta.$
\end{definition}

\begin{lemma}\label{pan6} (\cite[Theorem 1]{ern1}, \cite[Theorem 4.7]{ZH}).
For a poset $\mathcal{P}$, the following two conditions are equivalent:
\begin{enumerate}
    \item $\mathcal{P}$ is precontinuous;
    \item $\delta(\mathcal{P})$ is a continuous lattice.
\end{enumerate}
\end{lemma}

\begin{definition}\label{ast1}
Let \( (X, \tau) \) be a topological space and \( R \subseteq X \times X \) a binary relation.
We say that the relation \( R \) is \textit{Nachbin closed} (\cite{nac})  if the set
\vspace{5pt}
\begin{center}
$G(\preceq_{_R}) = \{ (x, y) \in X \times X \mid x \preceq_{_R} y \}$
\end{center}
is a closed subset of the product space \( (X \times X, \tau \times \tau) \); that is,
$(X \times X) \setminus G(\preceq_{_R}) \in \tau \times \tau.
$
Equivalently, \( R \) is Nachbin closed if for every sequence \( (x_n, y_n) \in G(\preceq_{_R}) \) such that \( x_n \to x \) and \( y_n \to y \), it follows that
$(x, y) \in G(\preceq_{_R})$, that is, $x \preceq_{_R} y$.
Also, by \cite[Page 26]{nac}, a relation is Nachbin closed with respect to the topology $\tau$ if and only if for every pair $x, y \in X$ such that $(x,y)\notin R$, there exist a decreasing open neighborhood $O_y$ of $y$ and an increasing open neighborhood $O_x$ of $x$ such that $O_x \cap O_y = \emptyset$.
\end{definition}

\begin{proposition}\label{pan9}
Let  $(X, R)$ be an abstract system such that $\mathcal{L} =(X,\preceq_{_R})$ is a continuous lattice. Then, $R$ is Nachbin closed in the Lawson topology $\lambda(\mathcal{L})$.
\end{proposition}

\begin{proof}
By \cite[Theorems III-1.9 and III-1.10]{GH}, $\sigma(\mathcal{L}) \vee \omega(\mathcal{L}) = \lambda(\mathcal{L})$ is a compact topology. To show that $\preceq_{_R}$ is closed in $\lambda(\mathcal{L})$, suppose that $(x,y)\notin \preceq_{_R}$ for some $x, y \in X$. By the remark of (\cite[Definition I-1.6]{GH}), there exists $z \in X$ such that $z \ll x$ and $(z,y)\in \preceq_{_R}$. Then, $X \setminus \{y\in X\vert z \preceq_{_R} y\}$ is a decreasing $\omega$-open neighborhood of $y$ (hence, Lawson) and 
$\{w\in X\vert z\ll w\}$
is an increasing Scott-open neighborhood of $x$ (hence, Lawson) such that 
$X \setminus \{y\in X\vert z \preceq_{_R} y\}\cap \{w\in X\vert z\ll w\}= \emptyset$. It follows 
from Definition \ref{ast1}
that $\preceq_{_R}$ is a $\lambda(\mathcal{L})\times \lambda(\mathcal{L})$-closed subset of $X \times X$. \end{proof}
By \cite[Theorem 1-6.4]{GW} we have that $\Delta$ is closed in  $\lambda(\mathcal{L})$

\begin{proposition}\label{pan21}
Let $\mathcal{L}=(X, \preceq)$ be a poset equipped with the Lawson topology $\lambda(\mathcal{L})$. Then, $(X,\lambda(\mathcal{L}))$ satisfies the $T_1$-order separation property.
\end{proposition}

\begin{proof}
Let $x, y \in X$ with $y \succ x$. Consider the set
\[
U = X \setminus \uparrow y = \{ z \in X : z \not\succeq y \}.
\]
This set is open in the lower topology (because it is the complement of a principal filter $\uparrow y$), and hence is also open in the Lawson topology (which contains the lower topology). Since $x \prec y$, it follows that $x \not\succeq y$, so $x \in U$. On the other hand,
$y \notin U$ (since $y \succeq y$), and all elements $z \succeq y$ are not in $U$.

Now let $(x_a)_{a \in A}$ be any net converging to $x$ in the Lawson topology. By definition of convergence, for any open neighborhood $U$ of $x$, there exists some $a_0 \in A$ such that for all $a \geq a_0$, $x_a \in U$. Therefore, for all such $a$, $x_a \not\succeq y$, i.e., $y \not\preceq x_a$, which in the strict order means $y \succ x_a$. Thus, there exists an $a_0$ so that $y \succ x_a$ for all $a \geq a_0$.

Therefore, both conditions of the $T_1$-order separation property are satisfied in the Lawson topology.
\end{proof}

\begin{proposition}\label{pan11}
Let \( (X, \preceq) \) be a poset and let $\mathcal{M}(X,\preceq)$
is non-empty and stable.
Then:
\[
X = \bigcup_{m \in \mathcal{M}(X)} [m, \to), \quad \text{where } [m, \to) = \{ x \in X \mid m \succeq x \},
\]
and each such set \( [m, \to) \) is maximal with respect to inclusion among \( \preceq \)-chains, totally ordered, and upward-directed.
\end{proposition}

\begin{proof}
Fix any \( x \in X \). Since \( \mathcal{M}(X) \) is stable with respect to $\preceq$, there exists \( m_0 \in \mathcal{M}(X,\preceq) \) such that \( m_0 \succeq x \). If \( x \notin \mathcal{M}(X) \), then \( x \) is not maximal, so there exists \( x_1 \in X \setminus \mathcal{M}(X,\preceq) \) such that \( x_1 \succeq x \). Since \( \mathcal{M}(X,\preceq) \) is stable, there exists \( m_1 \in \mathcal{M}(X,\preceq) \) such that \( m_1 \succeq x_1 \), and thus:
\[
m_1 \succeq x_1 \succeq x.
\]
We can repeat this process: for each \( x_{_{k-1}} \in X \setminus \mathcal{M}(X,\preceq) \), there exists \( x_{k} \in X \) such that \( x_{k} \succeq x_{_{k-1}} \), and eventually some \( m_k \in \mathcal{M}(X,\preceq) \) appears such that:
\[
m_k \succeq x_{_{k}} \succeq \cdots \succeq x.
\]

If the process does not terminate in finitely many steps, we construct a transfinite sequence \( f : \alpha \to X \) indexed by 
ordinals \(( \alpha \in \mathrm{Ord})\), defined as:
\begin{itemize}
    \item \( f(0) = x \),
    \item If \( f(\alpha) \notin \mathcal{M}(X) \), choose \( f(\alpha+1) \in X \setminus \mathcal{M}(X,\preceq) \) such that \( f(\alpha+1) \mathrel{R} f(\alpha) \),
    \item For limit ordinals \( \alpha \), as it is well-known, an ordinal \( \alpha \) is a limit ordinal if and only if there exists an ordinal \( \alpha^{\prime} < \alpha \), and for every \( \alpha^{\prime} < \alpha \), there exists an ordinal 
 $\alpha^{\prime\prime}$ such that \( \alpha^{\prime}<\alpha^{\prime\prime} < \alpha \).
 \end{itemize}   
Therefore, by repeating the steps above for \( \alpha, \alpha^{\prime}, \alpha^{\prime\prime}\),
instead of \( f(0), f(1), f(2), \dots \), and considering them with the same layout, we conclude that:
\[
f(\alpha^{\prime\prime})\succeq \, f(\alpha^{\prime}) \quad \text{and} \quad f(\alpha^{\prime}) \ne f(\alpha^{\prime}) \quad \text{for all } \alpha^{\prime}, \alpha^{\prime\prime} \in \mathrm{Ord} \text{ satisfying } 
\alpha^{\prime}<\alpha^{\prime\prime}<\alpha.
 \]
Therefore, \( f(\alpha) \) holds. It follows that \( f(\alpha) \) is true for all \( a \in \mathrm{Ord} \).

Since \( \mathcal{M}(X,\preceq) \) is stable, the process must terminate at some ordinal \( a_{_\infty} \), with \( f(a_{_\infty}) \in X\setminus \mathcal{M}(X,\preceq) \). We then have a chain:
\[
m\succeq f(a_{_\infty}) \succeq  \cdots \succeq  f(0) = x.
\]

Now, define \( C_x \subseteq X \) as the image of the sequence \( f \), i.e., the set of all points that lie on the \( \preceq \)-chain from \( m  \) down to \( x \). Then \( C_x \) is totally ordered by \( \preceq \) and each step respects \( \preceq \).

Let \( \mathcal{F}_m \) be the family of \( \preceq \)-chains that:
\begin{itemize}
    \item contain \( m \in \mathcal{M}(X,\preceq) \),
    \item are totally ordered with respect to \( \preceq \),
    \item and contain all elements in \( C_x \).
\end{itemize}

The family \( \mathcal{F}_m \) is partially ordered by inclusion, and every chain in \( \mathcal{F}_m \) has an upper bound (the union of the chains), which is again totally ordered. Thus, by Zorn’s Lemma, there exists a maximal chain \( [m, \to) \in \mathcal{F}_m \).

Therefore, for every \( x \in X \), there exists \( m \in \mathcal{M}(X,\preceq) \) such that \( x \in [m, \to) \), and:
\[
X = \bigcup_{m \in \mathcal{M}(X,\preceq)} [m, \to).
\]

Finally, each \( [m, \to) \) is:
\begin{itemize}
    \item totally ordered by construction,
    \item upward-directed because any two elements \( x, y \in [m, \to) \) have a common upper bound (since it's a chain),
    \item and maximal with respect to inclusion among \( \preceq \)-chains by Zorn's Lemma.
\end{itemize}
\end{proof}

\begin{remark}
In our setting, the maximal elements are assumed not only to exist but also to form a stable set. This additional property is decisive: it ensures that the order-theoretic decomposition of the poset into upward-directed, totally ordered cones rooted at the maximal elements respects and reflects the original partial order.
In particular, although Zorn’s Lemma may justify the existence of a total ordering within each cone, the union of such cones, organized by stable maximal elements, defines a structure in which:
\begin{itemize}
\item
The original partial order remains unchanged throughout the construction,
\item
No artificial comparabilities or extensions of the order are introduced at any stage,
\item
The precontinuity condition is verified with respect to the initial order $\preceq$, not with respect to a modified or completion-based ordering.
\end{itemize} 
\end{remark}

\begin{proposition}\label{pan12}
Let \( (X, \preceq) \) be a poset and suppose that:
\begin{itemize}
    \item[(i)] The set of maximal elements \( \mathcal{M}(X,\preceq) \subseteq X \) is non-empty and stable with respect to $\preceq$.
    \item[(ii)] For each \( m \in \mathcal{M}(X,\preceq) \), the set \( [m, \to) = \{ x \in X \mid m \succeq x \} \) is totally ordered, upward-directed, and maximal with respect to set inclusion, and 
  \item[(iii)]  
    \[
X = \bigcup_{m \in \mathcal{M}(X)} [m, \to).
\]
\end{itemize}
Then, the poset \((X,\preceq) \) is precontinuous. 
\end{proposition}
\begin{proof}
Fix \( x \in X \). By condition (iii), there exists \( m \in \mathcal{M}(X, \preceq) \) such that \( x \in [m, \to) \). By (ii), the set \( [m, \to) \) is totally ordered and upward-directed. By (i), we may fix such an \( m \) for every \( x \in X \).

We define the way-below set of \( x \) as:
\[
\ll x := \{ a \in X \mid a \ll x \} := \bigcap \{ I \in \mathrm{Fid}(X) \mid x \in I^\delta \},
\]
where \( \mathrm{Fid}(X) \) is the collection of Frink ideals of \( X \), and \( I^\delta := (I^\uparrow)^\downarrow \). Notice that if \( I \) has a join, then \( x \in I^\delta \) implies \( x \leq \bigvee I \).

We will prove that \( x \in (\ll x)^\delta \). Let \( I \in \mathrm{Fid}(X) \) such that \( \ll x \subseteq I \). We want to show that \( x \in I^\delta \), i.e., every finite subset of \( \ll x \) has an upper bound in \( I \).

Since \( x \in [m, \to) \), consider any finite subset \( F \subseteq \ll x \cap [m, \to) \). Then:
\begin{itemize}
    \item \( F \subseteq \ll x \Rightarrow F \subseteq I \),
    \item \( F \subseteq [m, \to) \), which is upward-directed and totally ordered by (ii),
\end{itemize}
so \( F \) has an upper bound \( y \in [m, \to) \). Since \( I \) is downward closed and \( F \subseteq I \), the upper bound \( y \in I \).

Thus, every finite \( F \subseteq \ll x \) has an upper bound in \( I \), and hence \( x \in I^\delta \) by definition of Frink closure.

Since this is true for every \( I \in \mathrm{Fid}(X) \) with \( \ll x \subseteq I \), we conclude:
\[
x \in (\ll x)^\delta.
\]
Hence, the precontinuity condition is satisfied for every \( x \in X \), and so \( (X, \preceq) \) is precontinuous.
\end{proof}

As we said above, the stability of the maximal set ensures a canonical decomposition that supports precontinuity intrinsically in the given poset, independent of the use of external ordering extensions. This distinguishes our framework from those relying purely on Zorn’s lemma for the existence of order-theoretic structures and gives a stronger, order-preserving foundation to the result.

\begin{theorem}\label{1a1}
Let \( (X, R) \) be an abstract decision problem, where \( R \) is an acyclic binary relation on \( X \). The following statements are equivalent:
\begin{enumerate}
    \item[$(\mathfrak{a})$] The set of \( R \)-maximal elements, \(\mathcal{M}(X, R)\), is non-empty and stable under \( R \).
        \item[$(\mathfrak{b})$] There exists a compact topology \( \tau \) on \( X \) such that
        \begin{enumerate}
            \item[(i)] \( (X, \tau) \) satisfies the $T_1$-order separation property with respect to \( R \);
            \item[(ii)] the relation \( R \) is Nachbin closed.
        \end{enumerate}
\end{enumerate}
\end{theorem}
\begin{proof}($\mathfrak{a}$) $\Rightarrow$ ($\mathfrak{b}$) 
Suppose that \( \mathcal{M}(X, R) \) is a non-empty stable subset of \( X \) with respect to the relation \( R \). By Proposition~\ref{pan1}, the set \( \mathcal{M}(X, R) \) is also stable in the poset \( (X, \preceq_{_R}) \). By Propositions~\ref{pan11} and~\ref{pan12}, the poset \( (X, \preceq_{_R}) \) is precontinuous. Then, by Lemma~\ref{pan6}, the MacNeille completion \( (X, \preceq_{_R})^\delta \) is a continuous lattice.
Let \( \lambda(\mathcal{L}^\delta) \) denote the Lawson topology induced by the continuous lattice \( (X, \preceq_{_R})^\delta \). By Lemma~\ref{law}, \( \lambda(\mathcal{L}^\delta) \) is compact.
Furthermore, by Propositions~\ref{pan9} and~\ref{pan21}, the relation \( R \) is Nachbin closed and 
\( (X, \lambda(\mathcal{L}^\delta) \) satisfies the $T_1$-order separation property with respect to \( R \).

\par\smallskip\par\noindent
 ($\mathfrak{b}$) $\Rightarrow$ ($\mathfrak{a}$) 
 Let \((X, \tau)\) be a compact topological space, and let \(R \) be a binary relation 
 which
is Nachbin closed and 
\( (X, \tau) \) satisfies the $T_1$-order separation property with respect to \( R \).
Since $R$ is acyclic, 
by \cite[Theorem]{ber}, we have that 
$\mathcal{M}(X,R)\neq \emptyset$. 
Since for each $x,y \in \mathcal{M}(X,R)$ there holds $(x,y)\notin R$, we conclude that internal stability of a stable set is satisfied.
To prove external stability of a stable set, suppose to the contrary that for each $x\in X\setminus \mathcal{M}(X,R)$ we have that $(m,x)\notin R$ whenever $m\in \mathcal{M}(X,R)$.
Let
\begin{center}
$A=\{y\in X\setminus \mathcal{M}(X,R)\vert (m,y)\notin R$ for all $ m\in \mathcal{M}(X,R)\}$.
\end{center}

We prove that $A=\emptyset$ and thus $\mathcal{M}(X,R)$ satisfies the external stability of stable sets. 
We first prove that
$A$ is a closed subset of $(X,\tau)$. 
Suppose that $t$ belongs to the closure of
$A$. Then, there exists a net $(t_{_\beta})_{_{\beta\in B}}$ in
$A$ with $t_{_\beta}\to t$. 
Since \( t_{_\beta} \in X\setminus \mathcal{M}(X,R) \), for each \( \beta \in B \), there exists \( s_{_\beta} \in X \) such that 
$(s_{_\beta}, t_{_\beta})\in R\subseteq \preceq_{_R}$.
By the compactness of \( X \), we may assume (by passing to a subnet if necessary) that \( s_{_\beta} \to s \in X \).
Now, since \( t_{_\beta} \to t \), \( s_{_\beta} \to s \), and \( (s_{_\beta}, t_{_\beta})\in \preceq_{_R} \) for all \( \beta \), the Nachbin closedness of \( R \) implies that
$(s, t)\in \preceq_{_R}$. By Proposition \ref{pan1} $\mathcal{M}(X,R)$ is also maximal and stable set for $\preceq_{_R}$.
Thus, \( t \in X\setminus \mathcal{M}(X,R) \). 
We have to show that $t\in A$. 
Suppose to the contrary that $t\notin A$. It follows that $(m^{\ast},t)\in R$ for some $m^{\ast}\in \mathcal{M}(X,R)$.
Since  $(X,\tau)$ has the
$T_1$-order separation property with respect to $\preceq_{_R}$, there exists 
$\beta^{\prime}\in B$ and $\beta\geq \beta^{\prime}$
such that $(m^{\ast},t_{_\beta})\in R$, which contradicts 
$t_{_\beta}\in X\setminus \mathcal{M}(X,R)$. This contradiction shows that $t\in A$ and thus $A$ is a closed subset of $(X,\tau)$.
It follows that
$A$ is compact. 
For each $x\in X\setminus \mathcal{M}(X,R)$ there exists  $y_{_x}\in X\setminus \mathcal{M}(X,R)$
such that $(y_{_x},x)\in R\subseteq \preceq_{_R}$.
Since  $(X,\tau)$ has the
$T_1$-order separation property with respect to $\preceq_{_R}$, there exists an open neighborhood $U_{y_{_x}}$ of $x$
which satisfies $y_{_x} \succ_{_R} x$ ($R$ acyclic).
Therefore, for each $x\in X\setminus \mathcal{M}(X,R)$,
the sets $U_{y_{_x}}\bigcap (X\setminus \mathcal{M}(X,R))$ are open neighbourhoods of $x$ in the
relative topology of $X\setminus \mathcal{M}(X,R)$. Hence,
\begin{center}
$X\setminus \mathcal{M}(X,R)=\displaystyle
\bigcup_{x\in X} [U_{y_{_x}}\bigcap (X\setminus \mathcal{M}(X,R))]$.
\end{center}
Since the space $X\setminus \mathcal{M}(X,R))$ is compact 
in the
relative topology, there exist $\{x_{_1}, \ldots, x_{_n}\}$ such that
\[
X\setminus \mathcal{M}(X,R) = \bigcup_{i \in \{1, \ldots, n\}} U_{y_{_{x_i}}}.
\]
Consider the finite set $\{y_{_{x_{_1}}}, \ldots, y_{_{x_{_n}}}\}$. Then, for each 
$x \in X\setminus \mathcal{M}(X,R)=\bigcup\{U_{y_{_{x_i}}} \mid i = 1, \ldots, n\}$, 
there exists $i \in \{1, \ldots, n\}$ such that $y_{_{x_{_i}}}\succeq_{_R} x$. 
Since $y_{_{x_{_1}}} \in X\setminus \mathcal{M}(X,R)$, it follows that 
$y_{_{x_{_i}}}\succ_{_R}y_{_{x_{_1}}}$ for some $i \in \{1, \ldots, n\}$. 
If $i=1$, then we have a contradiction. Otherwise, call this element 
$y_{_{x_{_2}}}$.
Then, we have $y_{_{x_{_2}}}\succ_{_R} y_{_{x_{_1}}}$.
Similarly, $y_{_{x_{_3}}}\succ_{_R} y_{_{x_{_2}}}\succ_{_R} y_{_{x_{_1}}}$.
As 
 $\{y_{_{x_{_1}}}, \ldots, y_{_{x_{_1}}}\}$ is finite, by an induction argument based on this logic, we obtain the existence of a 
 $\succ_{_R}$-cycle. This last conclusion contradicts the acyclicity of $\succ_{_R}$.
 This leads to a contradiction with the acyclicity of $R$.
Therefore, $A=\emptyset$, which completes the proof.

\end{proof}

We will proceed to characterize the most important variants of the concept of stable sets in the order of their appearance, where each variant aimed to address and improve upon the weaknesses of its predecessors. This methodical approach is necessary to demonstrate how the properties and interactions of these variants of stable sets contribute to a comprehensive solution to the weaknesses pointed out in the classical definition of stable sets by von Neumann and Morgenstern.

The following lemma, which follows directly from Lemma~\ref{a221}, the relevant definitions, Van Deemen’s contraction theorem (see~\cite[Theorem 4.7]{van1}), and the results of Andrikopoulos~\cite[Theorem 3.1]{and},~\cite{and3}, provides the foundation for a clearer understanding of the subsequent theorems and their proofs.

\begin{lemma} \label{panm}
Let $(X, R) \in \Omega(X)$ and $ \mu(\Xi,\widetilde{R})= \{X^*_1, X^*_2, \dots, X^*_m\}$.
Then,
\begin{enumerate}
\item
$V = \mathcal{S}_{ch}(X,R) \quad \text{if and only if} \quad 
V = \displaystyle\bigcup_{X^*_i \in \mu(\Xi,\widetilde{R})} X^*_i.$
\item
$V \in  \mathcal{GS}(X,R) \quad \text{if and only if} \quad 
V = \displaystyle\bigcup \{x_i \in X^*_i\vert X^*_i \in \mu(\Xi,\widetilde{R})$

\item
If $V\in \mathcal{SS}(X,R)$ then
$
V \cap X^*_i \neq \emptyset \quad \text{for any } X^*_i \in \mu(\Xi,\widetilde{R}).
$
\item
$V \in \mathcal{M}\mathcal{S}(X, R) \quad \text{if and only if} \quad 
V = \displaystyle\bigcup_{X^*_i \in V^*} X^*_i 
\quad \text{where } V^* \subseteq \mu(\Xi,\widetilde{R}).$
\item
$\mathcal{S}_{ch}(X,R)\subseteq \mathcal{U}\mathcal{T}(X,R)$.

\end{enumerate}
\end{lemma}

We begin by characterizing the socially stable set proposed by Delver and Monsuur \cite{DM} , as well as the $m$-stable set introduced by 
Peris and Subiza \cite{per}.

\begin{definition} Let $X$ be a set and let $F \subseteq X$. The \emph{excluded set topology on $X$ generated by $F$} is the topology
\[
\tau_{\mathrm{exc}} = \{ U \subseteq X \mid U \cap F = \varnothing \} \cup \{ X \}.
\]

In other words, the open sets are all subsets of $X$ that are disjoint from $F$, together with $X$ itself. Then, $X$ is compact under $\tau$ since every open cover of $X$ includes $X$ itself. Hence, $\{ X \}$ is always a finite subcover (see \cite[Page 48]{SS}).
\end{definition}

\begin{theorem}\label{a272}
Let $R$ be a binary relation on $X$. The following conditions are equivalent:\par
(a) The Duggan set of $R$ in $X$ is non-empty,
\par
(b) There exists a compact topology $\tau$ on $X$ such that 
$(X,\tau)$ satisfies the weak $T_1$-order separation property with respect to $\mathcal{T}$.
\end{theorem}

\begin{proof}
($\mathfrak{a}$) $\Rightarrow$ ($\mathfrak{b}$) Suppose that $\mathcal{U}\mathcal{T}(X,R)$
is a non-empty Duggan set of $R$ in $X$. 
Let $\tau$ be the excluded set topology in $X$ generated by $\mathcal{U}\mathcal{T}(X,R)$.
Then, $X$ is compact 
under $\tau$. 
On the other hand, for each $x\in X$ we have $\{y\in X\vert x\overline{\mathcal{T}} y\}
\bigcap \mathcal{U}\mathcal{T}(X,R)=\emptyset$.
Therefore, whenever $y\overline{\mathcal{T}} x$, ($\mathcal{T}$ is acyclic) $\{z\in X\vert y\overline{\mathcal{T}} z\}$ is an open neighborhood of $x$ with $y\notin \{z\in X\vert y\overline{\mathcal{T}} z\}$. Hence, $(X,\tau)$ satisfies the $T_1$-order separation property with respect to $\mathcal{T}$.
\smallskip\par\noindent
($\mathfrak{b}$) $\Rightarrow$ ($\mathfrak{a}$)
We have two cases to consider: (i) There exists $x^{\ast}\in X$ such that for each $y\in X$ there holds 
$(y,x^{\ast})\notin \mathcal{T}$; (ii)
For each $x\in X$ there exists $y\in X$ such that
$(y,x)\in \mathcal{T}$.
\par\noindent
{\it Case} (i). We have that $x^{\ast}\in \mathcal{U}\mathcal{T}(X,R)$ and thus $x^{\ast}$ belongs to the Duggan set
\par\smallskip\par\noindent
{\it Case} (ii). 
In this case, we have that
for every $x\in X$ there exists $y_{_x}\in X$ such that $y_{_x}\mathcal{T}x$.
Since  $(X,\tau)$ has the weak
$T_1$-order separation property with respect to $\mathcal{T}$, there exists an open neighborhood $O_{y_{_x}}$ of $x$
which satisfies $y_{_x} \mathcal{T} x$.
Therefore, for each $x\in X$,
the sets $O_{y_{_x}}$ are open neighbourhoods of $x$. Hence,
\begin{center}
$X=\displaystyle
\bigcup_{x\in X} O_{y_{_x}}$.
\end{center}
Since the space $X$ is compact, there exist $\{x_{_1}, \ldots, x_{_n}\}$ such that
\[
X = \bigcup_{i \in \{1, \ldots, n\}} O_{y_{_{x_i}}}.\]
Since $y_{_{x_{_1}}} \in X$, it follows that 
$y_{_{x_{_i}}}\mathcal{T} y_{_{x_{_1}}}$ for some $i \in \{1, \ldots, n\}$. 
If $i=1$, then we have a contradiction. Otherwise, call this element 
$y_{_{x_{_2}}}$.
Then, we have $y_{_{x_{_2}}}\mathcal{T} y_{_{x_{_1}}}$.
Similarly, $y_{_{x_{_3}}}\mathcal{T} y_{_{x_{_2}}}\mathcal{T} y_{_{x_{_1}}}$.
As 
 $\{y_{_{x_{_1}}}, \ldots, y_{_{x_{_1}}}\}$ is finite, by an induction argument based on this logic, we obtain the existence of a 
 $\mathcal{T}$-cycle. This last conclusion contradicts the acyclicity of $\mathcal{T}$.
 This leads to a contradiction with the acyclicity of $R$.
Therefore, the Duggan set of $R$ in $X$ is non-empty.
\end{proof}

We recall some definitions from~\cite[\S 4]{and4}, related to the Duggan set:

\begin{definition}
A pair of sets $(U^*, U)$ with $U^*, U \subseteq X$ is said to be a \textit{$P(R)$-undominated pair of sets} in $X$ if and only if

\begin{enumerate}[label=(\roman*)]
    \item $\emptyset \neq U^* \subseteq U$,
    \item for no $x \in U^*$ is there a $y \in X \setminus U$ such that $(y, x) \in P(R)$, and
    \item if $U$ is a non-singleton set, then, for each $y \in U$, there exists $x \in U^*$ such that $(x, y) \in \overline{P(R)}$.

  \end{enumerate}
\end{definition}

Let $\preceq$ be the partial order of set inclusion on all subsets of $X \times X$, i.e., if 
$(A, B), (C, D) \in \mathcal{P}(X) \times \mathcal{P}(X)$, where $\mathcal{P}(X)$ denotes the power set of $X$, then

\[
(A, B) \preceq (C, D) \text{ if and only if } A \times B \subseteq C \times D.
\]

\begin{definition}
An $P(R)$-undominated pair of sets $(U^*,U)$ is a \textit{minimal $P(R)$-undominated pair of sets} if there is no $P(R)$-undominated pair of sets $(F^*,F)$ satisfying $(F^*,F) \preceq (U^*,U)$. A non-empty subset $U^*$ of $X$ is called \textit{minimal $P(R)$-pairgenerator}, if there exists $U \subseteq X$ such that $(U^*,U)$ is a minimal $P(R)$-undominated pair of sets.
\end{definition}

\begin{definition}
A pair of sets $(U^*,U)$ is called \textit{pair $R$-cycle}, if $U^*,U$ are non-empty subsets of $X$ and they are $P(R)$-cycles in $X$. The pair $P(R)$-cycle $(U^*,U)$ is a \textit{top pair $P(R)$-cycle} if it is a minimal $P(R)$-undominated pair of sets. A non-empty subset $U^*$ in $X$ is a \textit{top $P(R)$-cycle pairgenerator}, if there exists $U \subseteq X$ such that $(U^*, U)$ is a top pair $P(R)$-cycle.
\end{definition}

\begin{lemma}\label{jes} (\cite[Theorem 12]{and4}).
Let \( (X, R) \) be an abstract decision problem. Then, the following sets are  equivalent: 
\begin{enumerate}[label=(\alph*)]
    \item \textit{the Duggan set},
    \item \textit{the union of all $P(R)$-undominated elements of $P(R)$ and all top $P(R)$-cycle pairgenerators in $X$},
    \item \textit{the set of maximal elements of $\mathcal{T}$}.
    \end{enumerate}
\end{lemma}

\begin{theorem} \label{a112}
Let \( (X, R) \) be an abstract decision problem. The following statements are equivalent:
\begin{enumerate}
    \item[$(\mathfrak{a})$] The family $\mathcal{SS}(X,R)$ of  socially stable sets of $(X,R)$ is non-empty.
        \item[$(\mathfrak{b})$] There exists a compact topology \( \tau \) on \( X \) such that
        \begin{enumerate}
            \item[(i)] \( (X, \tau) \) satisfies the $T_1$-order separation property with respect to \( \mathcal{T} \);
            \item[(ii)] the relation \( \mathcal{T} \) is Nachbin closed.
        \end{enumerate}
\end{enumerate}
\end{theorem}
\begin{proof}($\mathfrak{a}$) $\Rightarrow$ ($\mathfrak{b}$) 
Suppose that $\mathcal{SS}(X,R)$ is non-empty.
Define
\begin{center}
$\widetilde{\mathcal{S}} = \left\{ S \subseteq \mathcal{P}(X) \;\middle|\; \forall x, y \in S,\ \text{if } (x, y) \in \overline{P(R)} \big|_S, \text{ then}\  
(y,x) \in \overline{P(R)} \big|_S
\right\}$,
\end{center}
that is, \(S \in \widetilde{\mathcal{S}}\) satisfies the internal stability condition of socially stable sets. Then, $\widetilde{\mathcal{S}}$ is non empty since
$\mathcal{SS}(X,R)\in \widetilde{\mathcal{S}}$.
Take a chain \(\mathfrak{S}\) in  $\widetilde{\mathcal{S}}$ and let
$\widehat{S} = \bigcup \mathfrak{S}$.
Then, it is evident that $\widehat{S} $ belongs to $ \mathfrak{S}$.
Therefore, by Zorn's Lemma, $ \mathfrak{S}$ has an element, say $S_{_0}$, that is maximal with respect to set inclusion. 
We show that, for each \(x \in X\), it holds that \(\{ y \in X \mid x \mathcal{T} y \} \cap S_{_0} = \emptyset\), and therefore 
$S_{_0}$ is maximal set with respect to $\mathcal{T}$ (untraped set).
To prove it, we have two cases to consider: \ \ $x$ belongs to $S_{_0}$ or not.
If $x\in S_{_0}$, then since $S_{_0}$ is maximal in $\widetilde{\mathcal{S}}$ and $(x,y)\in P(R)\big|_{S_{0}}\subseteq \overline{P(R)}\big|_{S_{0}}$
we have that $(y,x)\in \overline{P(R)}\big|_{S_{0}}$. It follows that  \((x, y) \notin \mathcal{T}\).
If $x\notin S_{_0}$, then
suppose by contradiction that there exists
 \(y \in S_{_0}\) such that \((x, y) \in \mathcal{T} \subseteq P(R)\).  
Since \(\mathcal{SS}(X,R)\subseteq S_{_0}\), we have that \(x \notin \mathcal{SS}(X,R)\).  
Then, because of the external stability of socially stable sets, we have \((s, x) \in P(R)\) for some \(s \in \mathcal{SS}(X, R)\).  
It follows that \((s, y) \in \overline{P(R)}\).  
Since \(s, y \in \mathcal{SS}(X,R) \subseteq S_0\), we have \((y, s) \in \overline{P(R)}\).  
It follows that \((y, x) \in \overline{P(R)}\), which implies that \((x, y) \notin \mathcal{T}\).
Therefore,  $S_{_0}$ is the set of maximal elements in $X$ with respect to  $\mathcal{T}$, that is, $S_{_0}=\mathcal{M}(X,\mathcal{T})$.
On the other hand, $\mathcal{T}$ is acyclic (see \cite[p. 499]{dug}). 
Finally, due to the external stability property of socially stable sets, it follows that for each $x \in X \setminus \mathcal{SS}(X, R)$, there exists some $s \in \mathcal{SS}(X, R)$ with $(s, x) \in P(R) \subseteq \preceq_{\mathcal{T}}$.
Therefore, the conditions of the assumptions of Theorem \ref{1a1} are satisfied.
Then, similarly to the proof of implication ($\mathfrak{a}$) $\Rightarrow$ ($\mathfrak{b}$) of Theorem \ref{1a1}
there exists a compact topology $\tau$ on $X$such that 
the relation \( \mathcal{T} \) is Nachbin closed and 
\( (X, \tau) \) satisfies the $T_1$-order separation property with respect to \( \mathcal{T} \).

\vspace{5pt}
\par\noindent
($\mathfrak{b}$) $\Rightarrow$ ($\mathfrak{a}$). 
Let \((X, \tau)\) be a compact topological space, and let \(R \) be a binary relation 
 such that $\mathcal{T}$
is Nachbin closed and 
\( (X, \tau) \) satisfies the $T_1$-order separation property with respect to \( \mathcal{T} \).
Then, by Theorem \ref{a272}, we have that the Duggan set is non-empty.
By Lemma \ref{jes}, we have that $\mathcal{U}\mathcal{T}(X,R)$ is not empty. We prove that $\mathcal{U}\mathcal{T}(X,R)$
is a socially stable set. We first prove the internal stability of socially stable sets. Let $x, y\in \mathcal{U}\mathcal{T}(X,R)$ such that
$(x,y)\in \overline{P(R)}$, that is, 
$(x,y)\in \overline{P(R)}\big|_{\mathcal{U}\mathcal{T}(X,R)}$. By Lemma \ref{jes} we have that $x, y$ belong to
a top $P(R)$-cycle pairgenerator in $X$. Hence, $(y,x)\in \overline{P(R)}\big|_{\mathcal{U}\mathcal{T}(X,R)}$. To prove the external stability of socially stable sets, let
\begin{center}
$\mathcal{D} = \{ x \in X \setminus \mathcal{U}\mathcal{T}(X,R)\vert$\ for each \ $u\in \mathcal{U}\mathcal{T}(X,R)$\ we have \ $(u,x)\notin \mathcal{T}$\ \}.
\end{center}
Thus, by applying the implication (\(\mathfrak{b}\)~$\Rightarrow$~(\(\mathfrak{a}\)) of Theorem~\ref{a272} with $R$ replaced by $\mathcal{T}$, we deduce the existence of a $\mathcal{T}$-cycle in $\mathcal{D}$, which contradicts the acyclicity of $\mathcal{T}$ (see~\cite[p.~499]{dug}).

\end{proof}

We now characterize the $m$-stable set.
In finite cases, this set exists for every abstract decision problem. 
 Both the admissible set and the maximal alternative set are $m$-stable sets. In this sense, the notion of $m$-stable sets extends the notions of maximal alternative set and of the Schwartz 
 set. In general, an $m$-stable set solution is not core-inclusive (see \cite{per}, \cite{han}).

\begin{theorem}\label{a1123} Let $(X,R)$ be an abstract decision problem. Then, the following conditions are
equivalent: 
\par
($\mathfrak{a}$) 
The family $\mathcal{MS}(X,R)$ of $m$-stable sets of $(X,R)$ is non-empty.
\par
($\mathfrak{b}$) There exists a compact topology $\tau$ on $X$ such that
$(X,\tau)$ has the weak $T_1$-order separation property with respect to $R$.
\end{theorem}
\begin{proof}($\mathfrak{a}$) $\Rightarrow$ ($\mathfrak{b}$). 
Let $\mathcal{M}\mathcal{S}(X,R) \neq \emptyset$ be the family of $m$-stable sets of $(X,R)$. We first show that $\mathcal{S}_{ch}(X,R) \neq \emptyset$. To establish this, we prove that
\begin{center}
$\mathcal{M}\mathcal{S}(X,R)\bigcap \mathcal{S}_{ch}(X,R)\neq \emptyset.$
\end{center}
Suppose, for the sake of contradiction, that
\[
\mathcal{M}\mathcal{S}(X,R) \bigcap \mathcal{S}_{ch}(X,R) = \emptyset.
\]
Then, there exists an element $x_{_0} \in X$ such that $x_{_0} \in \mathcal{M}\mathcal{S}(X,R)$ but $x_{_0} \notin \mathcal{S}_{ch}(X,R)$.
By the definition of $\mathcal{S}_{ch}(X, R)$, there exists $y \in X$ such that $y \overline{P(R)} x_{_0}$. Since $\mathcal{M}\mathcal{S}(X,R)$ is externally stable, it follows that $y \in \mathcal{M}\cal{S}(X,R)$. By the internal stability of $m$-stable sets, we obtain $x_{_0} \overline{P(R)} y$.
Therefore, for every $y \in \mathcal{M}\mathcal{S}(X,R)$, we have $(y, x_{_0}) \notin P(\overline{P(R)})$. Moreover, for every $y \in X \setminus \mathcal{M}\mathcal{S}(X,R)$, we also have $(y, x_{_0}) \notin P(\overline{P(R)})$.
Hence, by \cite[Page 335]{and3}, we conclude that $x_{_0} \in \mathcal{S}_{ch}(X,R)$, which is a contradiction.
It follows that $\mathcal{S}_{ch}(X,R) \neq \emptyset$.
Let $ \tau_{\text{exc}}$ be the excluded set topology in $X$ generated by $\mathcal{S}_{ch}(X,R)$. Then, $ \tau_{\text{exc}}$ is compact.
On the other hand, for each $y\in X$, $\{x\in X\vert yP(\overline{P(R)})x\}\bigcap \mathcal{S}_{ch}(X,R)=\emptyset$. Hence, for each $y\in X$
the set
$\{x\in X\vert yP(\overline{P(R)})x\}\bigcap \mathcal{S}_{ch}(X,R)$ is an open neigborhood of $x$.
That is, whenever $y\succ_{_R}x$ , there exists an open neighborhood of $x$ which does not contain $y$.
Therefore, $(X,\tau_{\text{exc}})$ has the weak \emph{$T_1$-order separation property with respect to} $R$.

\par\smallskip\par\noindent
($\mathfrak{b}$) $\Rightarrow$ ($\mathfrak{a}$). 
We have two cases to consider: (i) There exists $x\in X$ such that for each $y\in X$ there holds 
$(y,x)\notin \overline{P(R) }$; (ii)
For each $x\in X$ there exists $y\in X$ such that
$(y,x)\in \overline{P(R)}$.
\par\noindent
Case (i). In this case, we have \( x \in \mathcal{S}_{ch}(X, R) \). Then \( x \) is either a \( P(R) \)-undominated element of \( X \) or a member of a $P(R)$-top cycle. In both subcases, the conditions of internal and external stability for \( m \)-stable sets are fulfilled. Therefore, the \( m \)-stable set is non-empty.
\par\smallskip\par\noindent
Case (ii). In this case, we assume that for every \( x \in X \), there exists \( y \in X \) such that \( (y,x)\in P(R)\subseteq \overline{P(R)} \). 
In this case, we have that
for every $x\in X$ there exists $y_{_x}\in X$ such that $y_{_x}\succ_{_R} x$.
Since  $(X,\tau)$ has the weak
$T_1$-order separation property with respect to $R$, there exists an open neighborhood $O_{y_{_x}}$ of $x$
which satisfies $y_{_x} \succ_{_R} x$.
Therefore, for each $x\in X$,
the sets $O_{y_{_x}}$ are open neighbourhoods of $x$. Hence,
\begin{center}
$X=\displaystyle
\bigcup_{x\in X} O_{y_{_x}}$.
\end{center}
Since the space $X$ is compact,
there exist $\{x_{_1}, \ldots, x_{_n}\}$ such that
\[
X = \bigcup_{i \in \{1, \ldots, n\}} O_{y_{_{x_i}}}.\]
Then,
as the implication ($\mathfrak{b}$) $\Rightarrow$ ($\mathfrak{a}$)
of Theorem \ref{a272}, we obtain the existence of a 
 $ \succ_{_R}$-cycle which contradicts the acyclicity of $ \succ_{_R}$.
 Therefore, $\mathcal{MS}(X,R)$ is non-empty.

\end{proof}

We now characterize 
the extended stable set which was proposed by Han,  Van Deemenn and Samsura \cite{HDS}.

\begin{lemma}\label{t1}
\rm Let $(X, R)$ be an abstract decision problem such that the Schwartz set $\mathcal{S}_{\mathrm{ch}}(X, R^{\widetilde{\omega}})$ is non-empty. Then,
\[
\mathcal{S}_{\mathrm{ch}}(X, R^{\widetilde{\omega}})\subseteq \mathcal{M}(X, R^{\widetilde{\omega}}).
\]
\end{lemma}

\begin{proof} 
Let $x \in \mathcal{S}_{\mathrm{ch}}(X, R^{\widetilde{\omega}})$. Then, for each $y \in X$, we have:
\[
(y, x) \notin P\left( \overline{P(R^{\widetilde{\omega}})} \right) \quad \text{(see \cite{and3})}.
\]
Therefore, to prove that $\mathcal{S}_{\mathrm{ch}}(X, R^{\widetilde{\omega}})\subseteq \mathcal{M}(X, R^{\widetilde{\omega}})$ we must show that 
$P(R^{\widetilde{\omega}}) \subseteq P\left( \overline{P(R^{\widetilde{\omega}})} \right)
$.
Assume that $(y, x) \in P(R^{\widetilde{\omega}})$ and $(y, x) \notin P(\overline{P(R^{\widetilde{\omega}}}))$. Then,
$
(y, x) \in \overline{P(R^{\widetilde{\omega}})} \subseteq \overline{P(R)}$
and
$(x, y) \in \overline{P(R^{\widetilde{\omega}})} \subseteq \overline{P(R)}.
$
It follows that
$(x, y) \in R^{\widetilde{\omega}}$ and $(y, x) \in R^{\widetilde{\omega}}.
$
Thus,
$
(x, y) \notin P(R^{\widetilde{\omega}})$,
which is a contradiction. This contradiction shows that
$\mathcal{S}_{\mathrm{ch}}(X, R^{\widetilde{\omega}})\subseteq \mathcal{M}(X, R^{\widetilde{\omega}}).$
\end{proof}

\begin{lemma}\label{t2}
\rm Let $(X, R)$ be an abstract decision problem. Then,
the extended dominance relation $R^{\widetilde{\omega}}$ is acyclic.
\end{lemma}
\begin{proof}
We prove this by contradiction. Suppose that $R^{\widetilde{\omega}}$ contains a cycle. Then there exists a sequence $x_1, x_2, \ldots, x_n \in X$ ($n \geq 2$) such that:
$$x_1 R^{\widetilde{\omega}} x_2, \quad x_2 R^{\widetilde{\omega}} x_3, \quad \ldots, \quad x_{n-1} R^{\widetilde{\omega}} x_n, \quad x_n R^{\widetilde{\omega}} x_1$$

According to the definition of an extended binary relation, for each relation $x_i R^{\widetilde{\omega}} x_{i+1}$ (where $x_{n+1} = x_1$), there exist $s_i, t_i \in X$ such that:
\begin{center}
$x_i \mathfrak{I} s_i$,\ \ \ 
$s_i R^{\widetilde{\omega}} t_i$,\ 
\ \ 
$t_i \mathfrak{I} x_{i+1}$. 
\end{center}
Thus, the elements $x_i,\ i \in \{1, 2, \dots, n\}$, constitute a $P(R)$-cycle, which implies $(x_i, x_j) \notin R^{\widetilde{\omega}}$, yielding a contradiction. It follows that $R^{\widetilde{\omega}}$ is acyclic.
\end{proof}

\begin{theorem} \label{a117}
Let \( (X, R) \) be an abstract decision problem. The following statements are equivalent:
\begin{enumerate}
    \item[$(\mathfrak{a})$] There is a non-empty extended stable set of $R$ in $X$.
        \item[$(\mathfrak{b})$] There exists a compact topology \( \tau \) on \( X \) such that
        \begin{enumerate}
            \item[(i)]  \( (X, \tau) \) satisfies the $T_1$-order separation property with respect to \( R^{^{\widetilde{\omega}}} \);
            \item[(ii)] the relation \( R^{^{\widetilde{\omega}}} \) is Nachbin closed.
        \end{enumerate}
\end{enumerate}
\end{theorem}

\begin{proof}($\mathfrak{a}$) $\Rightarrow$ ($\mathfrak{b}$) . Let $\mathcal{E}=(\varepsilon_{_i})_{_{i\in I}}\neq \emptyset$ be the extended stable set of $R$.
We show that $\mathcal{E} \subseteq \mathcal{S}_{\mathrm{ch}}(X, R^{\widetilde{\omega}})$, and thus, by Lemma~\ref{t1}, we conclude that 
$\mathcal{E} \subseteq \mathcal{M}(X, R^{\widetilde{\omega}})$
(we recall that $\mathcal{S}_{\mathrm{ch}}(X, R)$ is
the union of all $R^{\widetilde{\omega}}$-undominated
elements and all top $P(R^{\widetilde{\omega}})$-cycles in $X$).
We continue by proving that $\mathcal{E} \subseteq \mathcal{S}_{\mathrm{ch}}(X, R)$. To do so, we begin with the case where $\mathcal{S}_{\mathrm{ch}}(X, R) \neq \emptyset$. The case where $\mathcal{S}_{\mathrm{ch}}(X, R) = \emptyset$ will be examined afterward.
We proceed by way of contradiction. 
Let $\varepsilon_{_i}\in \mathcal{E}, i\in I$ be such that $\varepsilon_{_i}\notin \mathcal{S}_{\mathrm{ch}}(X, R^{\widetilde{\omega}})$. Then, 
there exists \( y_i \in X \) such that
$y_i \,\overline{P(R^{\widetilde{\omega}})}\, \varepsilon_i.$ 
Since, by Lemma \ref{t2}, we know that $R^{\widetilde{\omega}}$ is acyclic, it follows that $y_i \overline{R^{\widetilde{\omega}}} \varepsilon_i.$
If
$y_{_i}\notin \mathcal{E}$, because of
external stability of domination, there exists $\varepsilon_{_{i^{\prime}}}\in \mathcal{E}$, for some $i^{\prime}\in I$, such that 
$\varepsilon_{_{i^{\prime}}}R^{^{\widetilde{\omega}}}y_{_i}$ which implies $\varepsilon_{_{i^{\prime}}}\overline{R^{\widetilde{\omega}}}
y_{_i}$. 
Hence, 
\begin{center}
$\varepsilon_{_{i^{\prime}}}\overline{R^{\widetilde{\omega}}}\varepsilon_{_i}$.
\end{center}
We have to cases to consider, when $i^{\prime}=i$ or not.
\par\noindent
{\it Case 1.}
If $i=i^{\prime}$, then 
\(\varepsilon_i \) is part of an \( \overline{R^{\widetilde{\omega}}} \)-cycle that contains \( \varepsilon_i \).
By Zorn's Lemma, consider the family of all \( \overline{R^{\widetilde{\omega}}} \)-cycles \( (\widehat{\mathcal{C}}_\gamma)_{\gamma \in \Gamma} \), where each \( \widehat{\mathcal{C}}_\gamma \subseteq X \) contains \( \{ \varepsilon_i \} \). 
By the maximality obtained from Zorn’s Lemma, $\varepsilon_i$ belongs to an $ R^{\widetilde{\omega}}$-top cycle, $\widehat{\mathcal{C}}{\gamma^{\ast}}$. Therefore, $\varepsilon_i \in \mathcal{S}_{\mathrm{ch}}(X, R)
$. 
\par\noindent
{\it Case 2.} If $i\neq i^{\prime}$, following this process, there exists $i^{\prime\prime}\in I$ such that 
\begin{center}
$\varepsilon_{_{i^{\prime\prime}}}\in  \mathcal{S}_{\mathrm{ch}}(X, R)$\ \ ({\rm Case 1)}\ \ {\rm or}\ \
$\varepsilon_{_{i^{\prime\prime}}}\overline{R^{\widetilde{\omega}}}\varepsilon_{_{i^{\prime}}}\overline{R^{\widetilde{\omega}}}\varepsilon_i$,
\ 
$\varepsilon_{_{i^{\prime\prime}}}\overline{R^{\widetilde{\omega}}}\varepsilon_{_{i^{\prime}}}$, $\varepsilon_{_{i^{\prime\prime}}}\overline{R^{\widetilde{\omega}}}\varepsilon_{_{i}}$, \ $\varepsilon_{_{i^{\prime}}}\overline{R^{\widetilde{\omega}}}\varepsilon_{_{i}}$
\ {\rm and}\ $\varepsilon_{_{i^{\prime\prime}}}\notin  \mathcal{S}_{\mathrm{ch}}(X, R)$.
\end{center}
Hence,
\[
\varepsilon_{i} \,\overline{R^{\widetilde{\omega}}}\, \varepsilon_{j}, \quad
\varepsilon_{i} \neq \varepsilon_{j}, \quad
\text{for all } i, j \in \{0,1,2\} \text{ with } j < i,
\]
and
\[
\bigcup_{i \in \{1,2,3\}} \left( \{\varepsilon_{i}\} \cap \mathcal{S}_{\mathrm{ch}}(X, R) \right) = \varnothing.
\]
We refer to the above as the $T(0)$-property.
Let $I$ be an index set enumerating the above process.  
Without loss of generality, we assume that  
$\mathcal{I} = (I,\leq)$ is a well-ordered set; the existence of such an ordering follows from the Axiom of Choice.  
We proceed by transfinite induction on the well-ordered set $\mathcal{I}$.
According to transfinite induction, if $T(i)$ is true whenever $T(i^{\prime})$ is true for all $i^{\prime}<i$, then $T(i)$ is true for all $i$.

We proceed by transfinite induction on the well-ordered set $\mathcal{I} = (I,\leq)$, considering the following cases:

\begin{itemize}
    \item[\textbf{(A)}] \textit{Base case:} Show that the property $T(0)$ holds for the minimal element $0 \in I$.

    \item[\textbf{(B)}] \textit{Successor case:} For any successor ordinal $i+1$, prove that $T(i+1)$ follows from $T(i)$.

    \item[\textbf{(C)}] \textit{Limit case:} For any limit ordinal $i$, prove that $T(i)$ holds assuming $T(i')$ holds for all $i' < i$.
\end{itemize}
\par\smallskip
\par\noindent
{\it Step} (A). 
 As the base case of our transfinite induction, we adopt the $T(0)$-property established above.
\par\smallskip
\par\noindent
\textit{Step (B).} 
Let $j$ be a successor ordinal, and let $i$ be an ordinal such that $j < i$. 
If $i$ is a limit ordinal, then there exists an ordinal $\kappa^{\prime}$ with $j < \kappa^{\prime} < i$; 
otherwise, there exists an ordinal $\kappa$ such that $j < i < \kappa$. 
We prove that $T(j+1)$ follows from $T(j)$ in the case where $j < i < \kappa$ and $j+1 = \kappa$ 
(the case $j < \kappa^{\prime} < i$ with $j+1 = i$ is similar).
Repeating the arguments established above for 
$\varepsilon_{_{i}}, \varepsilon_{_{i^{\prime}}}, \varepsilon_{_{i^{\prime\prime}}}$
- now applied to the ordinals $j$, $i$, and $\kappa$ in the same configuration - 
we obtain
\[
\varepsilon_{_i} R^{^{\widetilde{\omega}}} \varepsilon_{_j}, \quad
\varepsilon_{_i} \neq \varepsilon_{_j}, \quad
\text{for all } i, j \in I \text{ with } j < i < \kappa .
\]
Therefore, $T(j+1) = T(\kappa)$ holds.

\par\smallskip
\noindent
\textit{Step (C).} 
Let $i^{\prime\prime}$ be a limit ordinal. 
By definition, $i^{\prime\prime}$ is a limit ordinal if it has no immediate predecessor and, 
for every $i < i^{\prime\prime}$, there exists $i^{\prime}$ such that $i<i^{\prime}<i^{\prime\prime}$.  
Assume as induction hypothesis that $T( i^{\prime})$ holds for all $ i^{\prime} <  i^{\prime\prime}$.  
Applying the same type of reasoning as above - but now directly to the ordinals 
$ i,  i^{\prime},  i^{\prime\prime}$ and the corresponding $\varepsilon_{i}, \varepsilon_{ i^{\prime}}, \varepsilon_{ i^{\prime\prime}}$ -
we deduce
\[
\varepsilon_{ i^{\prime\prime}} R^{^{\widetilde{\omega}}} \varepsilon_{ i^{\prime}}, \quad
\varepsilon_{ i^{\prime\prime}} \neq \varepsilon_{ i^{\prime}}, \quad
\text{for all } i^{\prime}, i^{\prime\prime} \in I \text{ with } i < i^{\prime} < i^{\prime\prime}.
\]
Therefore, $T(i)$ holds. Combining Steps~(A)–(C), it follows that $T(i)$ is valid for every $i \in I$ and, in particular,  
\[
\mathcal{E} \cap \mathcal{S}_{\mathrm{ch}}(X, R) = \varnothing.
\]
In this process, where the choice of starting element $\varepsilon_i$ plays a crucial role, we make the following key observations:
\begin{enumerate}
\item[($\mathfrak{A}$)]  
        For each index $k$, there exists some $\ell>k $ with $\varepsilon_\ell R^{^{\widetilde{\omega}}}\varepsilon_k $.\footnote{The choice 
        ``$\ell > k$'' is essential; it forces every element to be bounded above by an \emph{earlier} index.}
\item[($\mathfrak{B}$)] $\mathcal{E} \cap \mathcal{S}_{\mathrm{ch}}(X, R^{^{\widetilde{\omega}}}) = \varnothing.$
  \item[($\mathfrak{C}$)]  
        Any maximal chain that \emph{starts} with $\varepsilon_k$ we have $\varepsilon_\ell R^{^{\widetilde{\omega}}}\varepsilon_k $
        for any $\ell\neq k$.
   \end{enumerate}    
  Therefore, we have $\varepsilon_\ell R^{^{\widetilde{\omega}}} \varepsilon_k$ if we start from $k$ and $\varepsilon_k R^{^{\widetilde{\omega}}} \varepsilon_\ell$ if we start from $\ell$. This implies that $\varepsilon_k, \varepsilon_\ell$ belongs to a top $R^{^{\widetilde{\omega}}}$-cycle, that is $\varepsilon_k, \varepsilon_\ell$ belongs to $\mathcal{S}_{\mathrm{ch}}(X, R^{^{\widetilde{\omega}}})$, 
  a contradiction to ($\mathfrak{B}$). Hence, $\mathcal{E} \subseteq \mathcal{M}(X, R^{\widetilde{\omega}})$ which implies that 
  the set $\mathcal{M}(X, R^{\widetilde{\omega}})$ is a stable set with respect to $R$.
  
\par\smallskip\par\noindent
($\mathfrak{b}$) $\Rightarrow$ ($\mathfrak{a}$) . 
Let \((X, \tau)\) be a compact topological space, and let \(R \) be a binary relation 
 such that $R^{\widetilde{\omega}}$
is Nachbin closed and 
\( (X, \tau) \) satisfies the $T_1$-order separation property with respect to \( R^{\widetilde{\omega}}\).
Since $R^{\widetilde{\omega}}$ is acyclic, 
by \cite[Theorem]{ber}, we have that 
$\mathcal{M}(X,R^{\widetilde{\omega}})\neq \emptyset$. 
Since for each $x,y \in \mathcal{M}(X,R^{\widetilde{\omega}})$ there holds $(x,y)\notin R^{\widetilde{\omega}}$, we conclude that internal stability of an extended stable set is satisfied.
To prove external stability of a stable set, suppose to the contrary that for each $x\in X\setminus \mathcal{M}(X,R^{\widetilde{\omega}})$ we have that $(m,x)\notin R^{\widetilde{\omega}}$ whenever $m\in \mathcal{M}(X,R^{\widetilde{\omega}})$.
Let
\begin{center}
$\mathcal{H}=\{y\in X\setminus \mathcal{M}(X,R^{\widetilde{\omega}})\vert (m,y)\notin R^{\widetilde{\omega}}$ for all $ m\in \mathcal{M}(X,R^{\widetilde{\omega}})\}$.
\end{center}
Then, by applying the implication (\(\mathfrak{b}\)~$\Rightarrow$~(\(\mathfrak{a}\)) of Theorem~\ref{a272} with $R$ replaced by 
$R^{\widetilde{\omega}}$, we deduce the existence of a $R^{\widetilde{\omega}}$-cycle in $\mathcal{H}$, which contradicts the acyclicity of $R^{\widetilde{\omega}}$ (see Lemma \ref{t2}). Therefore, the external stability of an extended stable set is satisfied.
\end{proof}

The following lemmas will be employed to give a characterization of extended stable sets for irreflexive binary relations. These lemmas follow directly from the definition of an extended stable set and the concept of contraction.

\begin{lemma}\label{d121}{\rm Let $(X, R)$ be an abstract decision problem  and $(\Xi,\widetilde{R})$
be the contraction of it. Then, for any $X^{\ast}_{_i}, X^{\ast}_{_j}\in \Xi, (X^{\ast}_{_i}, X^{\ast}_{_j})\in \widetilde{R}$
if and only if $x_{_i}R^{^{\widetilde{\omega}}}x_j$ for all $x_{_i}\in X^{\ast}_{_i}$ and all $x_{_j}\in  X^{\ast}_{_j}$.
 }
\end{lemma}

The following theorem, which characterizes the solution of extended stable sets, is presented in \cite[Theorem 2]{HDS}.

\begin{theorem}
Let $(X, R)$ be an abstract decision problem where $X$ is finite and $R$ is asymmetric. 
Let 
\[
\mathfrak{E} \;=\; \{\mathcal{E}^{\ast}_{1}, \mathcal{E}^{\ast}_{2}, \dots, \mathcal{E}^{\ast}_{k}\}
\]
denote the family of extended stable sets of $\widetilde{R}$ on $\Xi$. 
Then $W \subseteq X$ belongs to the extended stable set of $R$ on $X$ 
if and only if 
\[
W = \{\varepsilon_1, \varepsilon_2, \ldots, \varepsilon_k\},
\]
where $\varepsilon_i \in \mathcal{E}^{\ast}_i$ for every $1 \leq i \leq k$.
\end{theorem}

In the finite case of the previous theorem, the non-emptiness of the extended stable set $\mathcal{E}$ is guaranteed whenever $R^{\widetilde{\omega}}$ is acyclic - a property which indeed holds. 
Thus, no additional assumptions are needed to ensure $\mathcal{E} \neq \varnothing$ in the finite setting. 
In contrast, in the infinite case, finiteness must be replaced by compactness, together with a suitable continuity condition on the relation $R$, in order to preserve the non-emptiness of $\mathcal{E}$.

\begin{theorem}\label{0987}
Let $(X,R)$ be an abstract decision problem, $\tau$ be a compact topology in $X$ such that 
\( (X, \tau) \) satisfies the $T_1$-order separation property with respect to \( R^{^{\widetilde{\omega}}} \)
and the relation \( R^{^{\widetilde{\omega}}} \) is Nachbin closed.
Suppose that
$\mathfrak{S}=(S_i^{\ast})_{_{i\in I}}$
is the stable set of $(\Xi,\widetilde{R})$. 
Then, 
$\mathcal{E}$ is an extended stable set of $(X,R)$ if and only if 
$\mathcal{E}=\{\varepsilon_{_i}\vert \varepsilon_{_i}\in S_i^{\ast}, i\in I\}$ is a set containing exactly one element from each 
$S_i^{\ast}$.
\end{theorem}
\begin{proof} 
Let $(X, \tau)$ be a compact topological space satisfying the $T_{1}$ order-separation property with respect to $R^{\widetilde{\omega}}$, and suppose that $R^{\widetilde{\omega}}$ is Nachbin closed.  
Let $\Xi$ be the quotient space obtained from $X$ by identifying points under the equivalence relation induced by $R$, and let 
\[
\pi : X \to \Xi
\]
denote the canonical projection. The relation $\widetilde{R}$ on $\Xi$ is defined as the natural projection of $R$ onto $\Xi$.

\medskip
\noindent
\textit{Step 1: Compactness is preserved under quotient maps.}  
The continuous image of a compact space is compact.  
Since $(X, R^{\widetilde{\omega}})$ is compact and $\widetilde{R}$ is induced by a continuous surjection from $R^{\widetilde{\omega}}$ (via the quotient map), it follows immediately that $(\Xi, \widetilde{R})$ is also compact.  
Hence, $(\Xi, \widetilde{R})$ inherits compactness from $(X, R^{\widetilde{\omega}})$.
\medskip
\par\noindent
\textit{Step 2: Preservation of the $T_{1}$ order-separation property.}  
Let $x^{\ast}_1, x^{\ast}_2 \in \Xi$ be two distinct elements such that $(x^{\ast}_2, x^{\ast}_1) \in \widetilde{R}$.  
Then there exist representatives $s \in x^{\ast}_2$ and $t \in x^{\ast}_1$ with $(s, t) \in P(R) \subseteq R^{\widetilde{\omega}}$.  
Since $(X, \tau)$ satisfies the $T_{1}$ order-separation property with respect to $R^{\widetilde{\omega}}$, there exists an open set $O \subseteq X$ such that $t \in O$ and $s \notin O$.  
Because the projection $\pi : X \to \Xi$ is continuous and open with respect to the quotient topology $\tau_{_\Xi}$, the set
\[
U = \pi(O)
\]
is open in $\Xi$, with $x^{\ast}_1 \in U$ and $x^{\ast}_2 \notin U$.  
Therefore, $(\Xi, \widetilde{R})$ also satisfies the $T_{1}$ order-separation property.

\medskip
\textit{Step 3: Preservation of Nachbin closedness.}  
The Nachbin closedness of $R^{\widetilde{\omega}}$ means that its graph is a closed subset of $X \times X$ endowed with $\tau \times \tau$.  
Since $\widetilde{R}$ is obtained as the image relation
\[
\mathrm{Gr}(\widetilde{R}) = (\pi \times \pi)\big( \mathrm{Gr}(R^{\widetilde{\omega}}) \big)
\]
under the continuous map $\pi \times \pi$, and $(X, \tau)$ is compact, the image of the compact set $\mathrm{Gr}(R^{\widetilde{\omega}})$ is compact and hence closed in the Hausdorff space $\Xi \times \Xi$.  
It follows that $\widetilde{R}$ is Nachbin closed.

Since $\widetilde{R}$ is acyclic, by Theorem \ref{1a1}, we have that the stable set $S(\Xi,\widetilde{R})=(S^{\ast}_i)_{_{i\in I}}$ 
of $\Xi$ with respect to $\widetilde{R}$
is non-empy.
Let $\mathcal{E}=\{\varepsilon_{_i} \vert \varepsilon_{_i}\in S_i^{\ast}, i\in I\}$.
For each $\varepsilon_{_j}, \varepsilon_{_i}\in \mathcal{E}$ there are distinct $S_j^{\ast}, S_i^{\ast}\in S(\Xi,\widetilde{R})$ such that $\varepsilon_{_j}\in S_j^{\ast}$ and $\varepsilon_{_i}\in S_i^{\ast}$.
Since $(S_j^{\ast},S_i^{\ast})\notin \widetilde{R}$, by Lemma \ref{d121},
we get that $(\varepsilon_{_j},\varepsilon_{_i})\notin R^{^{\widetilde{\omega}}}$. Hence, $\mathcal{E}$ satisfies the internal stability. 
To prove external stability of domination, let $z\in X\setminus \mathcal{E}$.
Then, there exists a $S_\kappa^{\ast}\in \Xi$ such that $z\in S_\kappa^{\ast}$.
If $S_\kappa^{\ast}\in S(\Xi,\widetilde{R})
$ there exists a $z^{\prime}\in S_\kappa^{\ast}$ such that 
$(\varepsilon_{_\kappa},z^{\prime})\in \mathfrak{I}$ and $(z^{\prime},z)\in P(R)$ implying $(\varepsilon_{_\kappa},z)\in R^{^{\widetilde{\omega}}}$.
Otherwise, $S_\kappa^{\ast}\notin S(\Xi,\widetilde{R})$. Then, there exists an $S_\lambda^{\ast}\in S(\Xi,\widetilde{R})$
such that $(S_\lambda^{\ast},S_\kappa^{\ast})\in \widetilde{R}$
with $\varepsilon_{_\kappa}\in S_\kappa^{\ast}$, $\varepsilon_{_\lambda}\in S_\lambda^{\ast}$
and $(\varepsilon_{_\lambda},z)\in R^{^{\widetilde{\omega}}}$. Hence, $\mathcal{E}$ satisfies the external stability. It follows that
$\mathcal{E}$ is an extended stable set.

Conversely, since $R^{^{\widetilde{\omega}}}$ is acyclic, by Theorem \ref{1a1}, we have that ${\mathcal{E}}\neq\emptyset$. Let
\begin{center}
$\mathcal{E}^{\ast}=\{X^{\ast}_i\in \Xi \vert x\in  \mathcal{E}, x\in X^{\ast}_i\}$.
\end{center}
Then, using Lemma \ref{d121} in a similar manner to the proof of Theorem 2 in \cite{HDS}, 
we conclude that $\mathcal{E}^{\ast}$ is an extended stable set of $(\Xi,\widetilde{R})$.

\end{proof}

We proceed to give a topological characterization of the existence of the $w$-stable set.

\begin{theorem}\label{a1} Let $(X,R)$ be be an abstract decision problem.
The following conditions are equivalent: 
\par
($\mathfrak{a}$) There is a non-empty $w$-stable set of $R$ in $X$.
\par
($\mathfrak{b}$) There exists a topology $\tau$ such that $(X,\tau)$
is compact and $(X,\tau)$ has the weak \emph{$T_1$-order separation property with respect to} $R$.
\end{theorem}
\begin{proof}($\mathfrak{a}$) $\Rightarrow$ ($\mathfrak{b}$) Suppose that $ \mathcal{W}\mathcal{S}(X,R)$
is non-empty. 
Let $ \tau_{\text{exc}}$ be the excluded set topology in $X$ generated by $\mathcal{W}\mathcal{S}(X,R)$. Then, $(X, \tau_{\text{exc}})$ is compact. It remains to prove that $(X,\tau_{\text{exc}})$ has the weak \emph{$T_1$-order separation property with respect to} $R$.
In fact, 
we first prove that for each $y\in X$ the sets $\{x\in  \mathcal{W}\mathcal{S}(X,R)\vert yP(\overline{P(R)})x\}$ are open in $ \tau_{\text{exc}}$. 
 We have two cases to consider: when $y\in  \mathcal{W}\mathcal{S}(X,R)$ or not.
 Let $y\in  \mathcal{W}\mathcal{S}(X,R)$, then because of internal stability of $w$-stable sets we have that $\{y\in \mathcal{W}\mathcal{S}(X,R)\vert yP(\overline{P(R)})x\}=\emptyset$. 
 If $x\notin  \mathcal{W}\mathcal{S}(X,R)$, then because of external stability of $w$-stable sets we also have  
 $\{y\in X\setminus \mathcal{W}\mathcal{S}(X,R)\vert yP(\overline{P(R)})x\}=\emptyset$ ($y\overline{P(R)}x$ implies $x\overline{P(R)}y$). 
 Therefore, whenever $yP(\overline{P(R)})x$, $O=\{y\in  \mathcal{W}\mathcal{S}(X,R)\vert xP(\overline{P(R)})y\}$
is an open neighborhood of $x$ with $y\notin O$. Hence, $(X,\tau)$ satisfies the weak $T_1$-order separation property with respect to $R$.
  \par\smallskip\par\noindent
 ($\mathfrak{b}$) $\Rightarrow$ ($\mathfrak{a}$) 
 Suppose that $\tau$ is compact 
 and $(X,\tau)$ has the weak \emph{$T_1$-order separation property with respect to} $R$.
 Suppose that $\tau$ is compact 
 and $(X,\tau)$ has the weak \emph{$T_1$-order separation property with respect to} $R$.
 We prove that $ \mathcal{W}\mathcal{S}(X,R)$
is non-empty.
Let $x^{\ast}\in X$. If $x^{\ast}$ is a $\overline{P(R)}$-maximal element, then $y\overline{P(R)}x^{\ast}$ implies 
$x^{\ast}\overline{P(R)}y$.
Hence, $\{x^{\ast}\}$ is a $w$-stable set.
Otherwise, $\mathcal{M}(X,{\overline{P(R)}})=\emptyset$. In this case,
for each $x\in X$ there exists $y_x\in X$ such that
$(y_x,x)\in P(\overline{P(R)})$.
Since  $(X,\tau)$ has the weak
$T_1$-order separation property with respect to $\mathcal{R}$, 
there exists an open neighborhood $O_{y_x}$ of $x$ such that $y_x \notin O_{y_x}$.
Therefore, for each $x\in X$,
the sets $O_{y_{_x}}$ are open neighbourhoods of $x$. Hence,
\begin{center}
$X=\displaystyle
\bigcup_{x\in X} O_{y_{_x}}$.
\end{center}
 Since the space $X$ is compact, there exist $\{x_{_1}, \ldots, x_{_n}\}$ such that
\[
X = \bigcup_{i \in \{1, \ldots, n\}} O_{y_{_{x_i}}}.\]
 Since $y_{_{x_{_1}}} \in X$, it follows that 
$y_{_{x_{_i}}}P(\overline{P(R)}) y_{_{x_{_1}}}$ for some $i \in \{1, \ldots, n\}$. 
If $i=1$, then we have a contradiction. Otherwise, call this element 
$y_{_{x_{_2}}}$.
Then, we have $y_{_{x_{_2}}}P(\overline{P(R)}) y_{_{x_{_1}}}$.
Similarly, $y_{_{x_{_3}}}P(\overline{P(R)}) y_{_{x_{_2}}}P(\overline{P(R)}) y_{_{x_{_1}}}$.
As 
 $\{y_{_{x_{_1}}}, \ldots, y_{_{x_{_n}}}\}$ is finite, by an induction argument based on this logic, we obtain the existence of a 
 $P(\overline{P(R)})$-cycle. 
 Let \( C_{\gamma_0} \) be the maximal \( P\big( \overline{P(R)} \big) \)-cycle containing \( C_{\gamma} \), whose existence is guaranteed by Zorn's Lemma. Then, 
$
\{ t_0 \}$, with $t_0 \in C_{\gamma_0},$
is a \( w \)-stable set.
The last conclusion completes the proof.

\end{proof}

The following theorem provides an alternative characterization of \( w \)-stable sets. This characterization is typically achieved via the contraction of \((X, P)\).
To proceed with this characterization, it is first necessary to generalize an important result, stated as Theorem 4.1 in~\cite{han}, which plays a key role in our analysis. In the finite case, the $w$-stable set is always nonempty; however, in the infinite case, this is not guaranteed. Therefore, the conditions given for the generalization of this result must ensure the nonemptiness of the $w$-stable set.

\begin{theorem}(\cite[Theorem 4.1]{han}).\label{papar}
Let $(X, R)$ be an abstract decision problem where $X$ is finite and $R$ is asymmetric. 
Let \[
\mu(\Xi,\widetilde{R})= \{X^{\ast}_{1}, X^{\ast}_{2}, \dots, X^{\ast}_{n}\}
\]
denote the set of maximal elements of $\widetilde{R}$ on $\Xi$. 
Then,
\begin{center}
 $W \in \mathcal{W}\mathcal{S}(X,R)$
if and only if 
$W \subseteq \{x_1, x_2, \ldots, x_n\}$,
\end{center}
where $x_i\in X^{\ast}_i$ and \ $1\leq i\leq n$.
\end{theorem}

\vspace{4mm}

\begin{theorem}\label{a121}{\rm 
Let $(X,R)$ be an abstract decision problem, and let $\tau$ be a compact topology in $X$.
Suppose that 
$(X, \tau)$ has the weak $T_1$-order separation property with respect to $R$.
 Then, 
 \begin{center}
$W$ is a non-empty $w$-stable set of $(X,R)$ if and only if 
$W\subseteq \{x_{_j}\vert j\in J\} $
\end{center}
where $ x_{_j}$ is exactly one alternative of 
$X^{\ast}_j\in \mu(\Xi,\widetilde{R}), J\subseteq I$.}
\end{theorem}
\begin{proof}
Let $(X,R)$ denote an abstract decision problem satisfying the assumptions of the theorem.
Let $W^{\ast}=\{X^{\ast}_{_j}\vert j\in J\}\subseteq \{X^{\ast}_i\vert i\in I\}$ be a subfamily of all ground sets which are $\widetilde{R}$-maximal 
in $\Xi$. This family is non-empty because of Lemma \ref{a221}. 
Let 
$x_{_j}\in X^{\ast}_j$ for each $j\in J$. We prove that $W=\{x_{_j}\vert j\in J\}$ is a 
$w$-stable set. If $(X,R)$ is strongly connected, then $X_i^{\ast}=X$ for all $i\in I$.
In this case, for each $x\in X$, $\{x\}$ is a $w$-stable set of $(X,R)$.
Suppose that
$X^{\ast}_{_{j^{\prime}}}\neq X^{\ast}_{_{j^{\prime\prime}}}$ for at least one pair 
$(j^{\prime},j^{\prime\prime})\in J\times J$.
The internal stability for $w$-set 
follows from \cite[Theorem 3.1]{and}.
To prove that $W$ satisfies external stability for $w$-set, let $x\in W$, $y\in X\setminus W$
such that $(y,x)\in \overline{P(R)}$. We have that $x\in X^{\ast}_{j^{\ast}}$ for some $j^{\ast}\in J$.

We have two cases to consider: when (1) $y\in \displaystyle\bigcup_{i\in I}X^{\ast}_{i}\setminus W$ and when 
(2) $y\in X\setminus \displaystyle\bigcup_{i\in I}X^{\ast}_{i}$.

In case (1), 
the only way for $(y,x)\in \overline{P(R)}$ to hold is for $y\in X^{\ast}_{j^{\ast}}$. But then, since $x\in X^{\ast}_{j^{\ast}}$
we have $(x,y)\in \overline{P(R)}$.

In case (2), since for each $i\in I$, $X_i^{\star}\in \mu(\Xi,\widetilde{R})$ we have that
$(y,x)\notin \overline{P(R)}$ for all 
$y\in X\setminus \displaystyle\bigcup_{i\in I}X^{\ast}_{i}$.
Therefore, this case does not exist. Hence,
$W$ satisfies external stability for $w$-set.

Conversely, let $W$ be a $w$-stable set.
By Theorem \ref{a1} we have that $W\neq \emptyset$.
Let $x_{_0}\in W$ and 
let $\{X^{\ast}_i\vert i\in I\}\subseteq \mu(\Xi,\widetilde{R})$.
We prove that $x_{_0}\in X^{\ast}_{\mathfrak{i}}$ for some $\mathfrak{i}\in I$.
If $x_{_0}$ is an $\overline{P(R)}$-maximal element, then $\{x_{_0}\}$ belongs to the Schwartz set, which implies that
$\{x_{_0}\}\in \mu(\Xi,\widetilde{R})$. Otherwise, there exists $y\in X$ such that $y\overline{P(R)}x_{_0}$.
By the external stability for $w$-sets it follows that $x_{_0}\overline{P(R)}y$.
By the Lemma of Zorn, 
the family of all $P(R)$-cycles
$(\widehat{\mathcal{C}}_{_\delta})_{_{\delta\in \Delta}}$,
$\widehat{\mathcal{C}}_{_\delta}\subseteq X$, which contain $\{x_{_0},y\}$ 
has a maximal element, let  $\widehat{\mathcal{C}}_{_{\delta_{_0}}}$. We prove that
$\widehat{\mathcal{C}}_{_{\delta_{_0}}}\in \mu(\Xi,\widetilde{R})$.
Indeed, let
$X^{\ast}\in \Xi$ such that $X^{\ast}\widetilde{R} \widehat{\mathcal{C}}_{_{\delta_{_0}}}$.
Then, there exists $t\in X^{\ast}$ and $s\in \widehat{\mathcal{C}}_{_{\delta_{_0}}}$ such that $(t,s)\in P(R)$. Since
$(s,x_{_0})\in \overline{P(R)}$, we conclude that $(t,x_{_0})\in \overline{P(R)}$.
By the external stability for $w$-sets we conclude that $(x_{_0},t)\in \overline{P(R)}$, which implies that 
$t\in \widehat{\mathcal{C}}_{_{\delta_{_0}}}$, which is impossible. 
Consequently, \( x_0 \in \widehat{\mathcal{C}}_{_{\delta_{_0}}} \), with \( \widehat{\mathcal{C}}_{_{\delta_{_0}}} \in \mu(\Xi, \widetilde{R}) \), which completes the proof.
\end{proof}

\section*{Acknowledgement}

\section*{Declarations}

{\bf Conflict of interest}\ The authors declare that there is no conflict of interest.

\par\bigskip\smallskip\par\noindent

\par\noindent
{\it Address}: {\tt {Athanasios Andrikopoulos} \\ {Department of Computer Engineering \& Informatics\\ University of Patras\\ Greece}
\par\noindent
{\it E-mail address}:{\tt aandriko@ceid.upatras.gr}

\par\noindent
{\it Address}: {\tt {Nikolaos Sampanis} \\ {Department of Computer Engineering \& Informatics\\ University of Patras\\ Greece}
\par\noindent
{\it E-mail address}:{\tt nsampanis.upatras.gr}

\end{document}